\documentclass[twocolumn]{aastex62}
\usepackage{import}

\usepackage{mathtools}
\usepackage{chngcntr}
\usepackage{footnote}
\usepackage[caption=false]{subfig}
\usepackage{float}
\usepackage{tabulary,booktabs}

\shorttitle{Large Scale Structures in the CANDELS Fields}
\shortauthors{Chartab et al.}

\begin{document}

\title{\textbf{\small Large Scale Structures in the CANDELS Fields: The Role of the Environment in Star Formation Activity}}

\correspondingauthor{Nima Chartab}
\email{nima.chartab@email.ucr.edu, n.ch.soltani@gmail.com}

\author[0000-0003-3691-937X]{Nima Chartab}

\affil{Department of Physics and Astronomy, University of California, Riverside, 900 University Ave, Riverside, CA 92521, USA}

\author{Bahram Mobasher}
\affiliation{Department of Physics and Astronomy, University of California, Riverside, 900 University Ave, Riverside, CA 92521, USA}

\author{Behnam Darvish}
\affiliation{Cahill Center for Astrophysics, California Institute of Technology, 1216 East California Boulevard, Pasadena, CA 91125, USA}

\author{Steve Finkelstein}
\affiliation{Department of Astronomy, The University of Texas at
Austin, Austin, TX 78712, USA}

\author{Yicheng Guo}
\affiliation{Department of Physics and Astronomy, University of Missouri, Columbia, MO 65211, USA}

\author{Dritan Kodra}
\affiliation{Department of Physics and Astronomy and PITT PACC,
University of Pittsburgh, Pittsburgh, PA 15260, USA}

\author{Kyoung-Soo Lee}
\affiliation{Department of Physics, Purdue University, 525 Northwestern Avenue, West Lafayette, USA}

\author{Jeffrey A. Newman}
\affiliation{Department of Physics and Astronomy and PITT PACC,
University of Pittsburgh, Pittsburgh, PA 15260, USA}

\author{Camilla Pacifici}
\affiliation{Space Telescope Science Institute, 3700 San Martin Drive, Baltimore, MD 21218, USA}

\author{Casey Papovich}
\affiliation{Department of Physics and Astronomy, Texas A\&M University, College Station, TX 77843-4242, USA}

\author{Zahra Sattari}
\affiliation{Department of Physics and Astronomy, University of California, Riverside, 900 University Ave, Riverside, CA 92521, USA}

\author{Abtin Shahidi}
\affiliation{Department of Physics and Astronomy, University of California, Riverside, 900 University Ave, Riverside, CA 92521, USA}

\author{Mark E. Dickinson}
\affiliation{National Optical Astronomy Observatories, 950 N Cherry
Avenue, Tucson, AZ 85719, USA}

\author{Sandra M. Faber}
\affiliation{University of California Observatories/Lick Observatory, University of California, Santa Cruz, CA 95064, USA}

\author{Henry C. Ferguson}
\affiliation{Space Telescope Science Institute, 3700 San Martin Drive, Baltimore, MD 21218, USA}

\author{Mauro Giavalisco}
\affiliation{Department of Astronomy, University of Massachusetts, 710 North Plesant Street, Amherst, MA 01003, USA}

\author{Marziye Jafariyazani}
\affiliation{Department of Physics and Astronomy, University of California, Riverside, 900 University Ave, Riverside, CA 92521, USA}

\begin{abstract}
 We present a robust method, weighted von Mises kernel density estimation, along with boundary correction to reconstruct the underlying number density field of galaxies. We apply this method to galaxies brighter than $\rm HST/F160w\le 26$ AB mag at the redshift range of $0.4\leq z \leq 5$ in the five CANDELS fields (GOODS-N, GOODS-S, EGS, UDS, and COSMOS). We then use these measurements to explore the environmental dependence of the star formation activity of galaxies. We find strong evidence of environmental quenching for massive galaxies ($\rm M \gtrsim 10^{11} \rm {M}_\odot$) out to $z\sim 3.5$ such that an over-dense environment hosts $\gtrsim 20\%$ more massive quiescent galaxies compared to an under-dense region. We also find that environmental quenching efficiency grows with stellar mass and reaches $\sim 60\%$ for massive galaxies at $z\sim 0.5$. The environmental quenching is also more efficient in comparison to the stellar mass quenching for low mass galaxies ($\rm M \lesssim 10^{10} \rm {M}_\odot$) at low and intermediate redshifts ($z\lesssim 1.2$). Our findings concur thoroughly with the "over-consumption" quenching model where the termination of cool gas accretion (cosmological starvation) happens in an over-dense environment and the galaxy starts to consume its remaining gas reservoir in depletion time. The depletion time depends on the stellar mass and could explain the evolution of environmental quenching efficiency with the stellar mass.
\end{abstract}


\section{Introduction}
It is now well established that the observed properties of galaxies are correlated with their host environment. In the local Universe, the environmental dependence of galaxy morphology  and star formation rate (SFR) confirms that early-type passive galaxies often reside in dense environments, such as galaxy groups and clusters, whereas late-type and star-forming systems are mostly found in less dense environments, so-called field \citep[e.g.,][]{Morphology,Kauffmann04,Balogh04,Peng10,Woo13}. However, the situation is not entirely settled at intermediate to high redshifts. While there is convincing evidence for a density-morphology relation at intermediate redshifts \citep[e.g.,][]{Capak07}, the exact trend in the density-SFR relation remains controversial. Some studies show a reversal relation so that on average the SFR is higher in dense environment \citep[][]{Elbaz07,Cooper08}, others find no significant correlation \citep[][]{Grutzbauch11,Scoville13,Darvish16} and some observe the same relation as in the local Universe \citep[][]{Patel2009}. Recently, an increasing number of studies have found that the locally observed environmental quenching persists at least out to $z\sim 2$ \citep[e.g.,][]{Fossati17,Guo2017,Kawinwanichakij2017,Ji18}. Therefore, a comprehensive study is needed to ascertain the role of the environment in star formation activity of galaxies at high redshifts.

Accurate measurement of the environment of galaxies is needed before any such study can be performed. One can use a variety of density estimators to quantify the environment in which galaxies are located. \citet{Darvish15a} have reviewed and compared different density estimators, including adaptive weighted kernel smoothing, $10^{th}$ and $5^{th}$ nearest neighbors, count-in-cell, weighted Voronoi tessellation and Delaunay triangulation. Comparing with simulations, they found that the weighted kernel smoothing method is more reliable than widely-used nearest neighbor and count-in-cell methods. Although kernel density estimation is a powerful and reliable technique for estimating the density field of galaxies, there are considerable ambiguities in the selection of the appropriate kernel function and optimized kernel window size (bandwidth). The selection of the bandwidth is the most crucial step in kernel density estimation. Small bandwidth results in an under-smoothed estimator, with high variability. On the other hand, large bandwidth causes an over-smoothed (biased) estimator. Boundary problem is another common issue regardless of the density estimator and the net effect is an underestimation of density near the boundaries since galaxies beyond the edge of the survey are missed. In this paper, we develop a new technique, weighted von Mises kernel density estimation considering boundary correction to reconstruct the density field of galaxies.

While measurement of density enhancement is available in contiguous wide-area surveys such as the Cosmic Evolution Survey (COSMOS) \citep{Scoville/COSMOS}, studying the influence of environment on the evolution of low mass galaxies ($\rm M \lesssim 10^{10} \rm {M}_\odot$) requires deep surveys that are often performed over much smaller areas because of the trade-off between the area coverage and the depth in galaxy surveys. The Cosmic Assembly Near-IR Deep Extragalactic Legacy Survey \citep[CANDELS;][]{2011ApJS..197...35G,2011ApJS..197...36K} includes extensive data in five fields, ideal for any study of the evolution of galaxies with redshift. The CANDELS provides: (1) Multi-waveband deep data from the Hubble Space Telescope (HST) and Spitzer Space Telescope for all the five fields; (2) Accurate measurement of the photometric redshifts, stellar mass and SFRs with their probability distributions; (3) Extensive spectroscopic observations for galaxies; (4) Constraints on the cosmic variance using five widely separated fields. These make the CANDELS fields ideal for such studies. The challenge, however, is to perform a reliable estimate of the density measurements for such fields with limited volume.

In this paper, we make a publicly available catalog of density measurements for 86,716 galaxies brighter than $\rm F160w\leq26 \ AB \ mag$ at $0.4 \leq z \leq 5$ in all the five CANDELS fields using weighted von Mises kernel density estimation with taking into account the boundary effect. We use a grid search cross-validation method to optimize the bandwidth of the kernel function. In order to reduce the projection effect, we use full photo-z probability distribution function (PDF) of individual galaxies (Kodra et al. in prep.).

The paper is organized as follows: In section \ref{sec:Data}, we discuss the data and describe the Spectral Energy Distribution (SED) fitting procedure to measure the physical properties of galaxies. Section \ref{sec:methodology} describes our methodology for measuring the local environment of galaxies and presents the galaxy environment catalog and large scale structure maps. In section \ref{sec:Result}, we explore the role of environment in the star formation activity of galaxies. We discuss our results in section \ref{sec:Discussion} and summarize them in section \ref{sec:Summary}.  

Throughout this work, we assume a flat
$\Lambda$CDM cosmology with $H_0=100h \rm \ kms^{-1} Mpc^{-1}$, $\Omega_{m_{0}}=0.3$ and $\Omega_{\Lambda_{0}}=0.7$. All magnitudes are expressed in the AB system and the physical parameters are measured assuming a Chabrier
IMF.

\section{Data}
\label{sec:Data}

\begin{figure}[b]
\centering
	\includegraphics[width=0.45\textwidth, clip=True, trim=0.1cm 1cm 0.7cm 2cm]{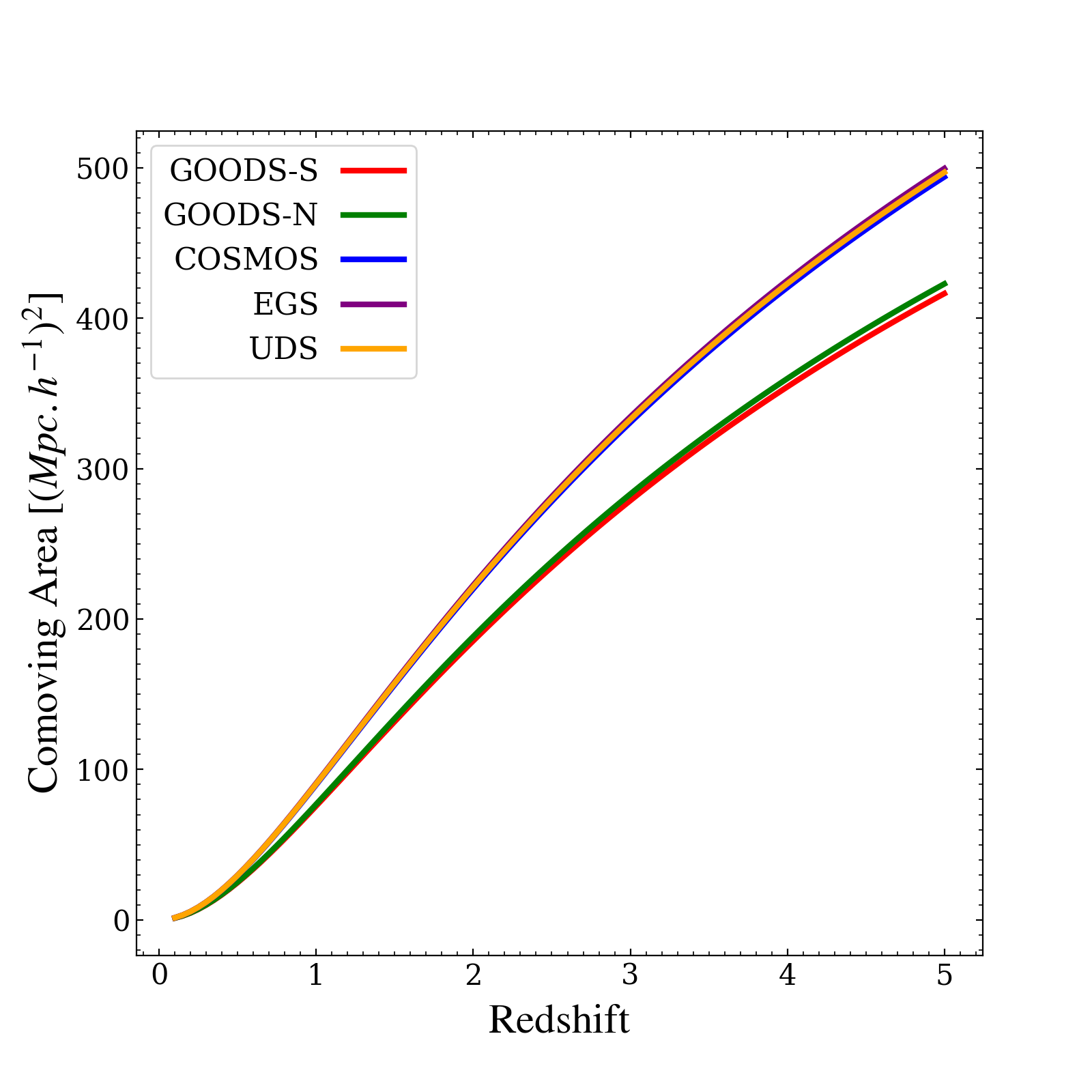}
	\caption{The comoving areal cover of the fields as a function of redshift. Each field covers $ \sim 25h^{-2}\ \rm  Mpc^2$ comoving area at $z\sim 0.5$ up to $\sim 450h^{-2}\ \rm Mpc^2$ at $z \sim 5$.}
	\label{Coverage}
\end{figure}

We use the HST/F160w (H-band) selected catalogs of the five CANDELS fields covering a total area of $\sim 960\ \rm arcmin^2$: GOODS-S \citep{Guo13}, GOODS-N \citep{Barro2019}, COSMOS \citep{Nayyeri17}, EGS \citep{Stefanon17}, and UDS \citep{Galametz13}. The comoving coverage area of each field as a function of
redshift is shown in figure \ref{Coverage}. 

The Catalogs are a combination of CANDELS wide, deep, and Hubble Ultra-Deep Field (HUDF) regions. The point source 5$\sigma$ limiting AB magnitude ranges from $\sim27.4$ to $\sim29.7$ in the wide and HUDF area, respectively. However, the 5$\sigma$ limiting magnitude is brighter for the extended objects and depends on the surface brightness profile of sources. The limiting magnitude in the wide field reaches $\rm H_{lim}\sim 26$, which corresponds to the 50\% completeness at the median size of sources \citep{Guo13}.

\begin{table*}

\caption{Summary of the data used in this work}
\centering 
\begin{tabular}{c c c c c c c} 
\hline\hline %inserts double horizontal lines
Field & Reference & Area ($\rm arcmin^2 $) & 5$\sigma$ depth (AB) & All objects & Final sample\footnote{Criteria: (1) $ \rm CLASS\_STAR < 0.9$, (2) Probability of being in $0.4 \le z \le 5$ greater than $ 95\%$, (3) H$\leq \rm 26\ AB\ mag$.} & spec/grism z\footnote{The percentage of galaxies in the final sample with confirmed spectroscopic/grism redshifts. } \\ [0.5ex] % inserts table
%heading
\hline % inserts single horizontal line
GOODS-S & \citet{Guo13} &   170 & 27.36& 34930 & 14200 & 16\%
\\
GOODS-N & \citet{Barro2019} &  170 & 27.8 & 35445 & 15746 & 18\%
\\
COSMOS &  \citet{Nayyeri17} &  216 & 27.56 & 38671 & 18896  & 7\%
\\
EGS &  \citet{Stefanon17} &  206 & 27.6 & 41457 & 19670 & 13\%
\\
UDS &  \citet{Galametz13} &  202 & 27.45 & 35932 & 18204 & 7\%
\\

\hline %inserts single line

\end{tabular}
\label{table:data}
\end{table*}

We utilize the updated CANDELS photometric redshift catalog (Kodra et al. in prep.), which provides accurate photometric redshifts with normalized median absolute deviation ($\rm \sigma_{NMAD}$) of $\sim 0.02$, combined with the spectroscopic/3D-HST grism redshifts ($z_{best}$). The catalog also contains redshift probability distribution functions (PDFs) of galaxies determined by the minimum Frechet distance method. The Frechet distance \citep{Alt1995} is a measure of similarity between two curves (e.g., two measurements of photo-z PDFs). The best PDF is obtained based on the minimum of the Frechet distance among six independent z-PDF measurements.  

In this work, we measure the local number density for a total of 86,716 galaxies selected based on the following criteria (Table \ref{table:data}):
\begin{itemize}
    \item Removing the stars by requiring SExtractor's stellarity parameter $ \rm CLASS\_STAR < 0.9$.
    \item Covering a redshift range of $0.4 \le z \le 5$. We select galaxies with greater than 95\% probability of being in this redshift range.  We limit our analysis to $z\geq 0.4$ due to the small volume of the survey at lower redshifts.
    \item A cut on H-band magnitude to remove the sources fainter than $\rm 26\ AB\ mag$. Although the fields have different $5 \sigma$ limiting magnitudes, we use a similar magnitude cut to have homogeneous and comparable samples.
\end{itemize}

\subsection{Stellar Mass \& Star Formation Rates}\label{SEDfitting}
We perform Spectral Energy Distribution (SED) fitting to derive physical parameters of galaxies such as stellar mass and star formation rate (SFR). We use the LePhare code \citep{Arnouts,Ilbert} combined with a library of synthetic spectra generated by the \citet{BC03} population synthesis code. To perform SED fitting, we fix redshifts on $z_{best}$ from the updated version of the CANDELS photometric redshift catalog. We assume an exponentially declining star formation history with nine e-folding times in the range of $0.01<\tau<30\ \text{Gyr}$. We adopt the  \citet{Chabrier} initial mass function, truncated at 0.1 and 100 $ \text{M}_{\odot}$, and \citet{Calzetti} attenuation law to apply dust extinction ($\rm E(B-V)\leq 1.1$). The code also includes emission lines using \citet{Kennicutt} relation between SFR and UV luminosity, as described in \citet{Ilbert2009}. Three different stellar metallicities are considered: $\rm Z=0.02, 0.008$, and 0.004. 

The LePhare code computes fluxes in all given bands for each template, then finds the template with minimum $\chi^2$ based on the model and observed fluxes. The best values of the physical parameters come from the template with the minimum $\chi^2$. The code also provides the median of the stellar mass (M), SFR and specific SFR (sSFR=SFR/M) along with uncertainties obtained from the marginalized probability distribution (probability$\rm \propto e^{-\chi^2/2}$) of each parameter. In this work, we use the median values for stellar mass, SFR and sSFR. We also obtain U,V and J rest-frame colors from best-fit SEDs.  

\begin{figure}[hb]
\includegraphics[width=0.5\textwidth,trim=0cm 1.1cm 0cm 1cm]{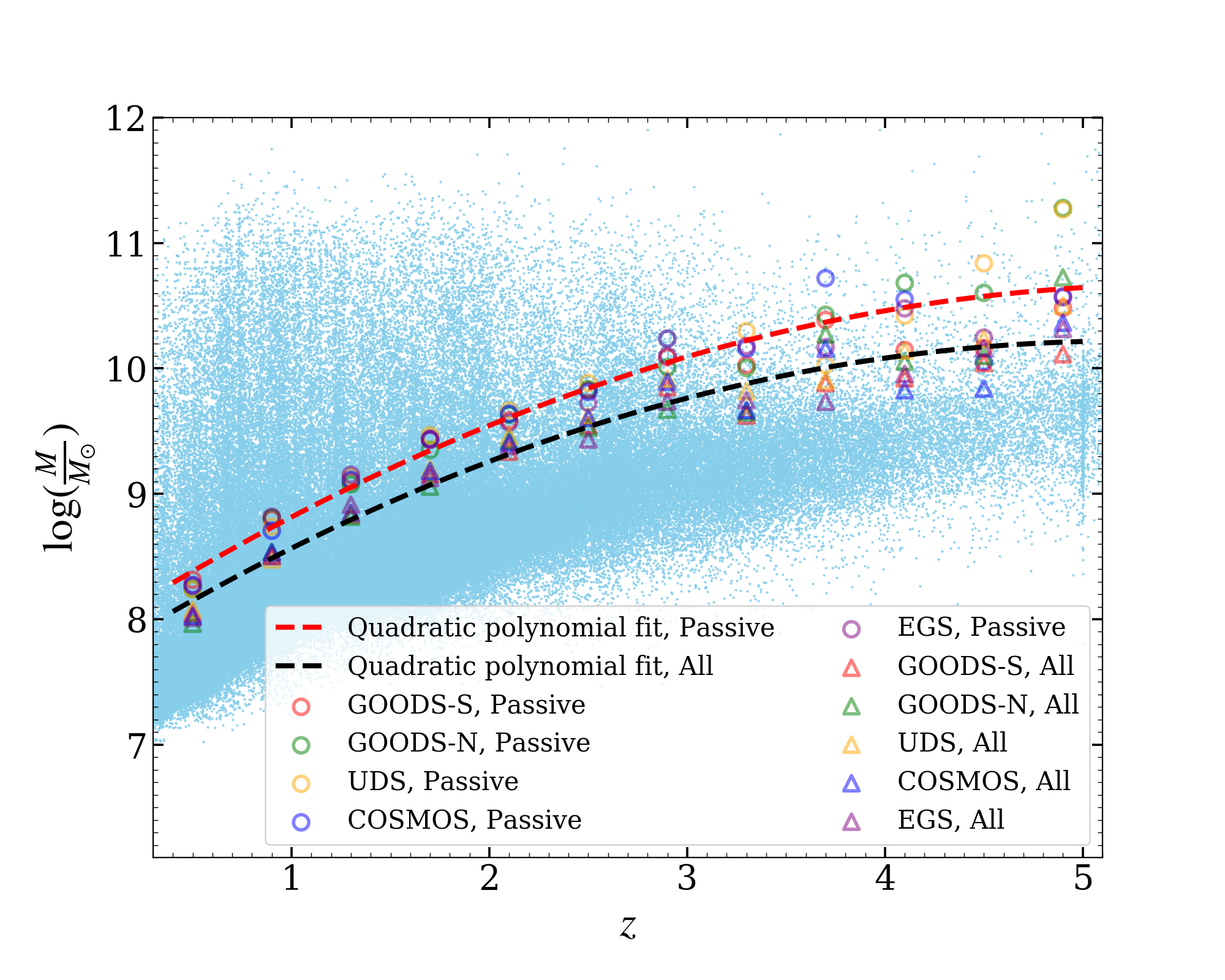}
\caption{The stellar mass of galaxies as a function of their redshifts. Red (black) dashed line represents the 95\% stellar mass completeness limit ($\rm H_{lim}=26$) for the passive (all) galaxies, determined using the method of \citet{Pozzetti}. }\label{Completness} 
\end{figure}

Figure \ref{Completness} shows the distribution of stellar mass as a function of redshift for galaxies in the five fields. The stellar mass completeness limit (95\%) associated with $\rm H_{lim}$=26 is determined using the method introduced by \citet{Pozzetti}. We divide the sample into redshift bins, separately in the case of all and quiescent populations. We utilize rest-frame U,V and J colors along with \citet{Muzzin12} criteria to select quiescent galaxies at $z < 4$. Beyond this redshift, we use a sSFR cut derived from the first quartile ($<25\%$ percentile) of the sSFR distribution to build a sub-sample of passive galaxies. We then measure the limiting stellar mass, $\rm M_{lim}$, for galaxies in the sub-sample, defined as the stellar mass a galaxy would have if it had a magnitude equal to the adopted magnitude limit of the survey ($\rm H_{lim}$). If we consider constant mass-to-light ratio, then $\rm M_{lim}$ for a galaxy with stellar mass $\rm M$ can be computed as $\rm \log M_{lim}=\log M+0.4(H-H_{lim})$. The stellar mass completeness limit ($\rm{ M_{min}}(z)$) is the 95th percentile of the $\rm M_{lim}$ distribution for the 20\% faintest sources at each redshift bin. Thus, if we take a sample of galaxies with the stellar mass higher than the completeness limit, less than 5\% of galaxies will be missed from the sample. As shown in figure \ref{Completness}, the stellar mass completeness limit is higher for passive galaxies with higher mass-to-light ratios. Hence, we adopt the completeness limit derived from passive galaxies, which can be modeled with a quadratic polynomial; $\log(\rm{M_{min}(z)/M_{\odot}})=7.90+z-0.09z^2$.      

\section{Measuring Galaxy Environment} \label{sec:methodology}
The environment of a galaxy is defined as the density field where that galaxy resides. To reconstruct the density field, we consider narrow redshift intervals (z-slices) and treat each z-slice as a two-dimensional structure. Using kernel density estimation, we calculate the density field within each z-slice. The location of a galaxy with photo-z is probabilistic and best identified by its redshift PDF. Thus, a galaxy with photometric redshift contributes to all z-slices. The contribution of each galaxy to different redshift intervals is proportional to the area under the photo-z PDF that lies within that interval. This introduces the weighted approach for density estimation. The density field associated with each galaxy is derived from the weighted sum of surface densities at different z-slices using the full redshift PDF of the galaxy. Therefore, the surface density, $\sigma$, of a galaxy at any given coordinate (RA, DEC) is,
\begin{equation}
\rm \sigma_{(RA,DEC)}=\sum_{j} \omega_{j} \sigma^{j}_{(RA,DEC)}
\label{eqn:surface_density}
\end{equation}
where $\rm \sigma^{j}_{(RA,DEC)}$ is the surface number density field at the position ($\rm RA$,$\rm DEC$) in the $j^{th}$ z-slice and $\omega_{j}$ is the probability of the desired galaxy to be in the $j^{th}$ z-slice. Although $\sim 12\%$ of our sample have spectroscopic/grism redshift, we do not use them to determine $\omega_{j}$. This assures that our method is not biased in favor of galaxies with accurate spectroscopic/grism redshifts. Therefore, we rely on uniformly calculated photometric redshifts, with well-calibrated probability distributions (Kodra et al. in prep.) in $\omega_{j}$ estimation. We use the area underneath the photometric redshift PDFs to obtain the likelihood of a galaxy to be in the $j^{th}$ z-slice.  

In order to measure $\rm \sigma^{j}_{(RA,DEC)}$, we use the weighted von Mises kernel density  estimation technique corrected for the boundary effect.  In the following sections, we describe different steps for estimating $\sigma^{j}$: an estimate of the redshift bin size (section \ref{RS}), weighted von Mises kernel density estimation (section \ref{sec:sigmaij}), bandwidth selection of the kernel function (section \ref{BS}) and boundary correction (section \ref{BC}).

\subsection{Selection of Redshift Slices}\label{RS}
\label{sec:redshift_slices}

It is important to optimize the width of redshift slices to account for the extended structures. While the redshift of a galaxy can be used to measure its location along the line of sight, the estimate can be affected by Redshift Space Distortion (RSD) due to the peculiar velocity of galaxies. The RSD effect is cosmological model dependent such that a galaxy cluster with internal velocity dispersion of $\Delta v$ will be extended in comoving space ($\Delta \chi$) as,
\begin{equation}
\Delta \chi=\frac{\Delta v}{H_{0}}\frac{(1+z)}{\sqrt{\Omega_{m_{0}}(1+z)^3+\Omega_{\Lambda_{0}}}}
\label{deltachi}
\end{equation}
where $\rm H_{0}, \Omega_{m_{0}}$ and $\rm \Omega_{\Lambda_{0}}$ are the present values of Hubble constant, matter density and dark energy density respectively. Hence, what we observe is the combination of the density and the velocity field. The proper binning in redshift space to reconstruct 2D maps of the large-scale structures is constrained by both the typical size of a galaxy cluster in redshift space and redshift uncertainties. In the presence of less accurate photometric redshifts, we have two options, either using a weighted scheme to incorporate the contribution of each galaxy in all z-slices accurately or adopting wide z-slices to collect all signals from galaxies with large redshift uncertainties. Here we use the weighted approach such that the width of z-slice is constrained based on the resolution of photo-z PDFs, $\Delta z/(1+z)\sim 1\%$ (Kodra et al. in prep.). This allows us to avoid over-smoothing caused by interlopers.   

The comoving size of a galaxy cluster due to the RSD effect (equation \ref{deltachi}), peaks at $z=(2\Omega_{\Lambda_{0}}/\Omega_{m_{0}})^{\frac{1}{3}}-1\simeq 0.65$. At that redshift, a massive galaxy cluster ($\Delta v\sim 1500\  \text{Km.s}^{-1}$) will be extended $\sim 18\ h^{-1}\text{Mpc}$ in comoving space due to the peculiar velocity of its galaxies. In addition, the estimated redshift uncertainty ($\Delta z/(1+z)\sim 1\%$) limits the z-slice width to $35\ h^{-1} \text{Mpc}$ ($\Delta v\sim 3000\  \text{Km.s}^{-1}$). Therefore, we fix the width of redshift bins (at all redshifts) to a constant comoving size of $35\ h^{-1} \text{Mpc}$ to satisfy both RSD effect and redshift uncertainty constraint. This results in a total of 124 z-slices spanning from $z=0.4$ to $5$. One should note that the constant comoving width does not imply a constant redshift interval. For comparison, the width of z-slice at $z=0.4$ is 0.014, while this value is 0.096 for $z=5$.

\subsection{Weighted von Mises Kernel Density Estimation}\label{sec:sigmaij}
The distribution of galaxies in each z-slice is analogous to a two-dimensional map where galaxies are labeled with their weights $\omega_{j}$, computed from the photometric redshift PDFs. These weights are assigned by determining the fraction of redshift PDF within each z-slice. A powerful non-parametric method for density estimation is weighted Kernel Density Estimation(wKDE)\citep{parzen1962} which can be written as:
\begin{equation}
\sigma^{j}(\mathrm{\textbf{X}_0})=\sum\limits_{i}\widetilde{\omega}_{i}^{j}\text{K}(\textbf{X}_i;\textbf{X}_0)   
\end{equation}
where $\sigma^{j}(\mathrm{\textbf{X}_0})$ is the estimated density at the position ${\textbf{X}_0}$ on $j^{th}$ z-slice and $\text{K}$ is the kernel function. The summation is over all data points ($\textbf{X}_i$) that exist in the desired z-slice. $\widetilde{\omega}_{i}^{j}$ is the normalized weight associated with $i^{th}$ data point, in the $j^{th}$ z-slice so that $\sum\limits_{i}\widetilde{\omega}_{i}^{j}=1$.

An appropriate choice of the kernel function for spherical data $\rm (RA,DEC)$ is the von Mises kernel \citep{Garcia13} expressed as, 
\begin{equation}
\text{K}(\textbf{X}_i;\textbf{X}_0)=\frac{1}{4\pi b^2 \sinh (1/b^2)} \exp(\frac{\cos \psi}{b^2})   
\end{equation}

where $b$ is the global bandwidth of the kernel function, which is the main parameter in the wKDE method and controls the smoothness of the estimate.  We will explain the bandwidth selection method in section \ref{BS}. $\psi$ is the angular distance between $\rm \textbf{X}_i=(RA_i,DEC_i)$ and $\rm \textbf{X}_0=(RA_0,DEC_0)$. $\cos \psi$ can be expressed as $\rm \sin{DEC_i}\sin{DEC_0}+\cos{DEC_i}\cos{DEC_0}\cos({RA_i-RA_0})$.

It should be noted that a Gaussian kernel function cannot be used in the case of spherical data. The kernel function must integrate to unity and a Gaussian function does not satisfy this requirement on the spherical space.

\subsection{Bandwidth selection}\label{BS}
\begin{figure}
	\includegraphics[width=0.5\textwidth]{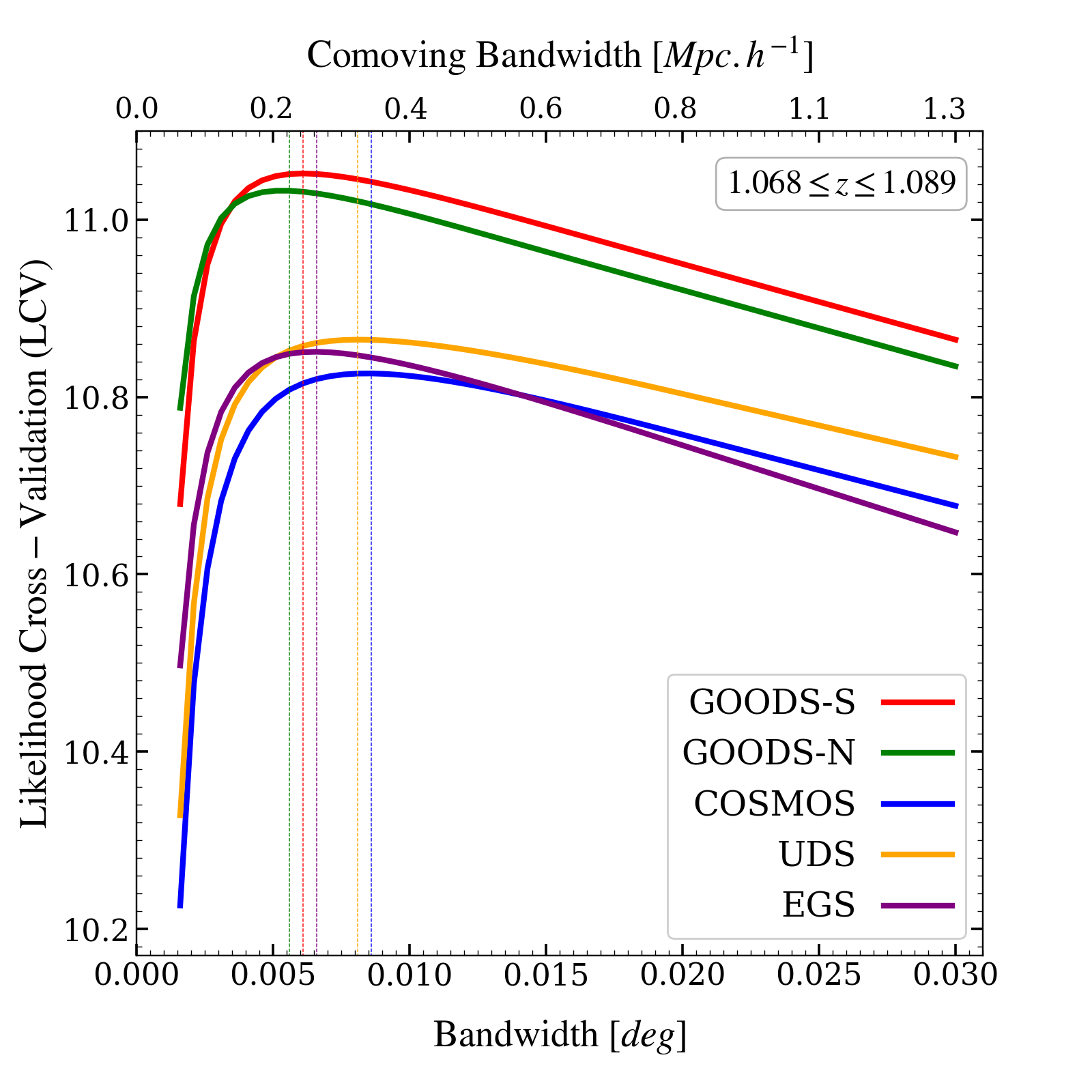}
	\caption{An example of Likelihood Cross-Validation optimization procedure at a given z-slice: $1.068 \leq z \leq 1.089$. We perform a grid search in the range of $ 0.0001^\circ \leq \mathrm{Bandwidth}\leq 0.03^\circ$ with 50 steps to maximize the LCV and find the best bandwidth (b). Dashed vertical lines show optimized bandwidths at $z\sim 1$.}
	\label{BWz1} 
\end{figure}

\begin{figure}[ht]
	\includegraphics[width=0.5\textwidth,clip=True, trim=0cm 5cm 0cm 5cm]{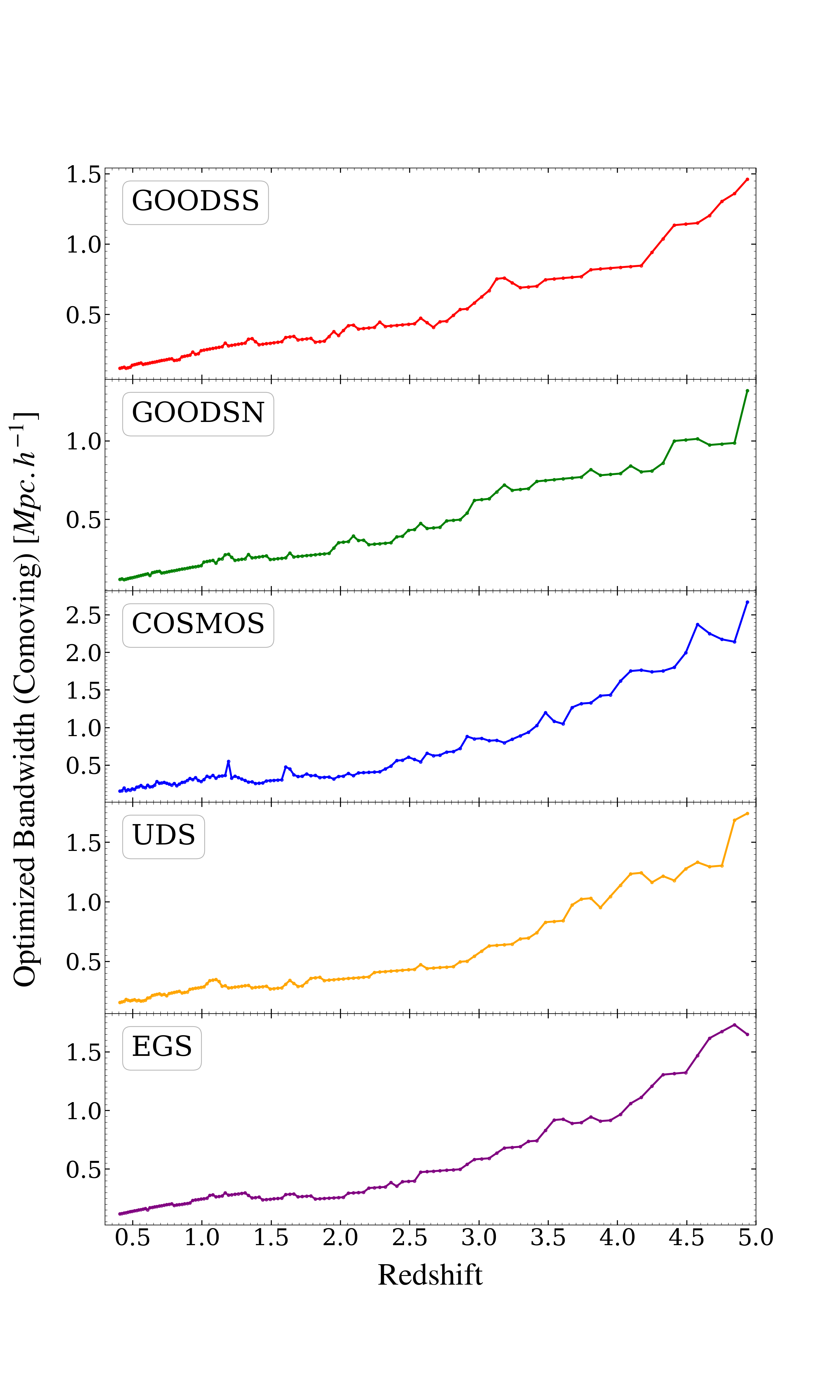}
	\caption{Optimized comoving global bandwidth (b) as a function of redshift for the five CANDELS fields. As we go to higher redshifts, we should increase the comoving size of the kernel function bandwidth to avoid the undesirable variance. }\label{BestBW} 
\end{figure}

Bandwidth selection is a challenging problem in kernel density estimation. Choosing too narrow bandwidth leads to a high-variance estimate (under-smoothing), while too wide bandwidth leads to a high-bias estimate (over-smoothing). This bias-variance trade-off can be solved by maximizing Likelihood Cross-Validation (LCV) \citep{Hall} which is defined by:
\begin{equation}
    \text{LCV(b)}=\frac{1}{N}\sum \limits_{k=1}^{N} \log \sigma_{-k}(\textbf{X}_k)
\end{equation}

where {$N$ is the total number of data points in a given z-slice and} $\sigma_{-k}(\textbf{X}_k)$ is the kernel estimator computed at position $\textbf{X}_k$ excluding the $k^{th}$ data point. We perform a grid search {in the range of $0.0001^\circ$ to $0.03^\circ$ with 50 steps} to find the optimized global bandwidth where $\rm LCV(b)$ is maximized. Figure \ref{BWz1} shows the LCV maximization results for one of the z-slices ($1.068 \leq z \leq 1.089$) in all CANDELS fields. For instance, the cross-validation method suggests $b=0.0061^\circ$ (a comoving distance of $0.26\ h^{-1} \text{Mpc}$) as the best bandwidth for the GOODS-S field at the mentioned z-slice. Figure \ref{BestBW} shows the optimized bandwidth in comoving coordinates for 124 redshift slices spanning from 0.4 to 5. 

A constant bandwidth ($b$) over each z-slice may result in under-smoothing in regions with sparse observations and over-smoothing in crowded areas. By varying the bandwidth for each data point ($i$) and defining a local bandwidth ($b_{i}$), we reduce the bias in dense regions and the variance in regions with sparse data. To incorporate adaptive smoothing, we vary the local bandwidth ($b_{i}$) as \citep{Abramson1982},
\begin{equation}
\label{adaptiveb}
b_{i}=b\Big\{\frac{\sigma(\mathrm{\textbf{X}_i})}{g}\Big\}^{-\alpha}
\end{equation}
where
{
\begin{align*}
\log g=\frac{1}{N}\sum \limits_{i=1}^{N} \log \sigma(\mathrm{\textbf{X}_i})
\end{align*}}
The sensitivity parameter, $\alpha$, is a constant which satisfies $0\le \alpha \le 1$ and can be fixed by simulation. In this study, we take a simple case where $\alpha=0.5$ as the sensitivity parameter does not have a significant effect in wKDE measurement \citep{Wang2007}. 
For each redshift slice, first, we estimate the bandwidth using the cross-validation method and then we employ the adaptive bandwidth technique (equation \ref{adaptiveb}) to reduce the variance/bias in the estimation. Finally, we need to correct the density field for the boundary effect, which is explained in the following section. 
\subsection{Boundary correction}\label{BC}
Kernel density estimation method assumes that the density field exists in the entire space. This assumption is not valid in most cases where a survey has data only for a small area of the sky. The trade-off between the area and the depth translates into a small coverage in deep surveys (e.g., CANDELS). Missing parts of the sky not covered in the survey result in an underestimation of the density field near the edge of the survey. Different methods have been developed to remove this problem, known as the boundary effect (e.g., Reflection method \citep{Schuster}, Boundary kernel method \citep{BCM} and Transformation method \citep{Trans}). Here we use the re-normalization method to correct for this boundary effect. 

The first order of the expectation value of the density fields can be written as \citep{Boundary}, 
\begin{equation}
\mathbb{E}(\sigma^{j}(\textbf{X}_0))\sim \sigma^{j}_{\text{True}}(\textbf{X}_0)\int_{\text{S}} \text{K}(\textbf{X}_i;\textbf{X}_0)
\end{equation}
where $\sigma^{j}_{\text{True}}$ is the true underlying density field and the integration is performed over the survey area ($S$). Thus, a simple way to correct the boundary is to re-normalize the density as,
\begin{equation}
\sigma^{j}_{\text{corr}}({\textbf{X}_0})={\sigma^{j}({\textbf{X}_0})}{n({\textbf{X}_0})}
\end{equation}
where $n({\textbf{X}_0})$ is the inverse of the kernel function integration centered at ${\textbf{X}_0}$ over the survey area ($S$),
\begin{equation}
n^{-1}({\textbf{X}_0})=\int_{\text{S}} \text{K}(\textbf{X}_i;\textbf{X}_0)
\end{equation}
This correction results in almost unbiased estimation of the density such that $\mathbb{E}(\sigma^{j}_{\text{corr}}({\textbf{X}_0}))\sim \sigma^{j}_{\text{True}}(\textbf{X}_0)$, but it may increase the variance close to the boundary.

\begin{figure}[ht]
	\includegraphics[width=0.48\textwidth,clip=True, trim=0.05cm 1.5cm 0cm 2cm]{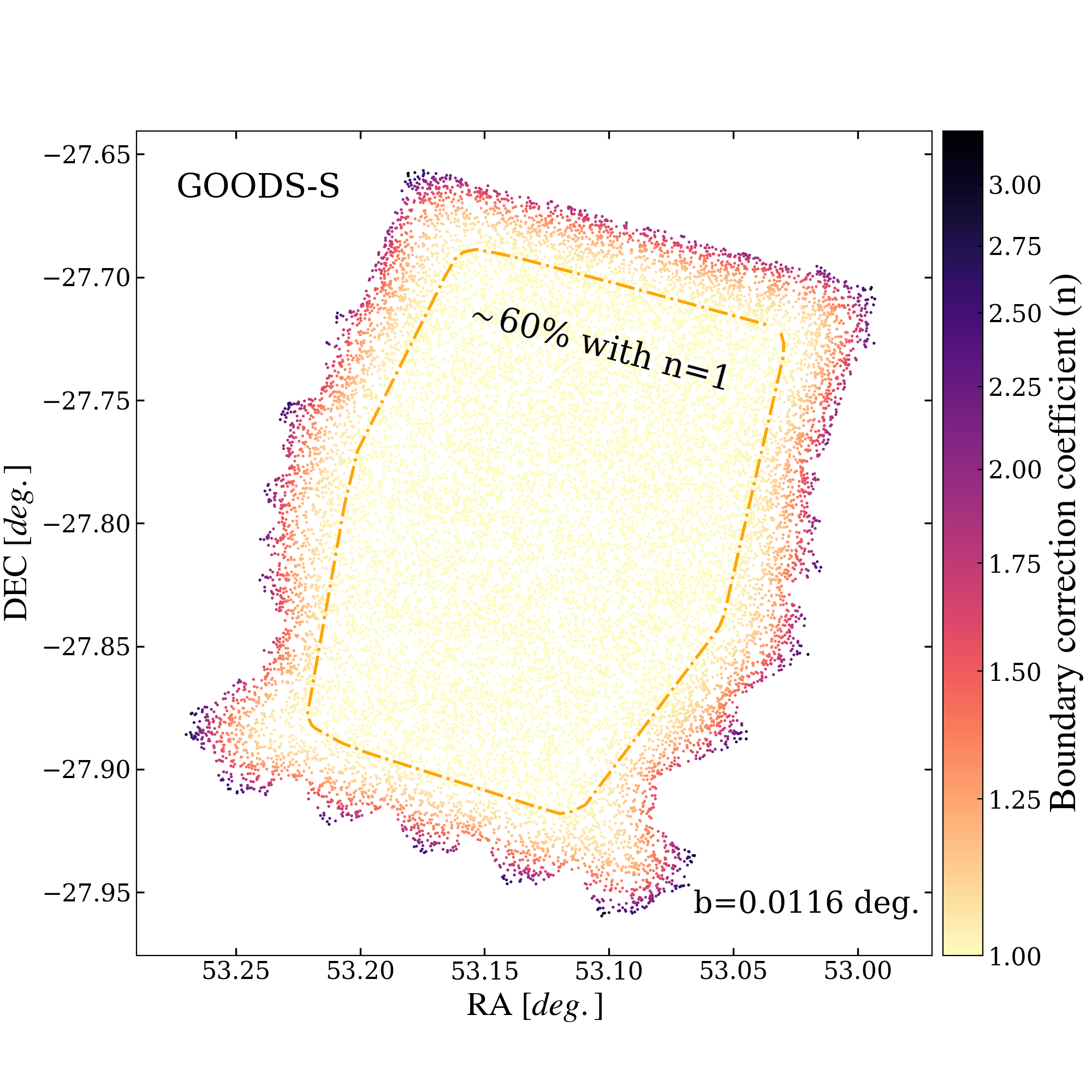}
	\caption{An example of the boundary correction coefficient (n) for galaxies in the GOODS-S field. It is shown that for the bandwidth of $0.0116^\circ$, the coefficient can be as large as $\sim 3.5$. Orange dash-dotted line encloses 60\% of galaxies which are not affected by boundary problem (n=1).}\label{EC} 
\end{figure}

Figure \ref{EC} shows the boundary correction coefficient (n) given the bandwidth of $0.0116^\circ$. At that bandwidth, 40\% of galaxies are affected by boundary problem ($n>1$) and it is crucial to apply boundary correction to those galaxies. Not correcting for boundary effects can result in an underestimation up to a factor of $\sim 3.5$.  

A common way to quantify the environment is defining density contrast ($\delta$) as,
\begin{equation}\label{DC}
    \delta=\frac{\sigma}{\Bar{\sigma}}-1
\end{equation}
where $\Bar{\sigma}$ is the background number density, which can be evaluated using $\sum\limits_{i} \omega^j_i/V$. Here, $\omega^j_i$ is the weight of $i^{th}$ galaxy at $j^{th}$ z-slice, and $V$ is the volume of corresponding z-slice. The boundary problem does not affect the background number density; however, it biases $\sigma$ close to the edge of the survey. Thus, boundary correction is necessary to avoid missing overdensities close to the edge.   

\subsection{Catalogs and density maps}
We utilize the \textit{boundary-corrected weighted von Mises kernel density estimation method} to reconstruct the density field of galaxies at $0.4 \leq z\leq 5$ in the five CANDELS fields. Details of the density measurement technique have been explained in sections \ref{RS}-\ref{BC} and summarized below:
\begin{itemize}
    \item Divide the survey into redshift slices with the comoving width of $\sim 35 h^{-1} \text{Mpc}$ (see section \ref{RS}).
\end{itemize}
For each z-slice:
\begin{itemize}
    \item Weight the galaxies using their redshift PDFs to construct the two-dimensional weighted maps (see section \ref{sec:methodology}).
    \item Perform a grid-search on the bandwidth space to minimize the LCV function and find the optimum bandwidths (see section \ref{BS}).
    \item Compute the density field associated with each galaxy using weighted KDE with the constant bandwidth drawn from the previous step (see section \ref{sec:sigmaij}) and apply the boundary correction technique (see section \ref{BC}).
    \item Make the bandwidth to be adaptive based on the boundary-corrected densities (see section \ref{BS}) and re-run the weighted KDE with the adaptive bandwidths. Then, reapply the boundary-correction method on the adaptively derived densities. 
\end{itemize}
The last step is to combine all z-slices to extract the density field associated with each galaxy.
\begin{itemize}
    \item For each galaxy, calculate the weighted summation of its density in all z-slices (see section \ref{sec:methodology}) to obtain the density field of the galaxy.
\end{itemize}

The full density field catalogs are available in the electronic version. Table \ref{table:examplecatalog} shows examples of the density measurements. The first four columns show the CANDELS ID, RA, DEC and redshift ($z_{best}$). The last three columns give the environmental properties, including comoving/physical density and density contrast. Comoving density is the number of galaxies in a cube with a comoving volume of $ 1\  h^{-3} \text{Mpc}^{3}$. The physical density can be computed by scaling the comoving density by a factor of $(1+z)^3$. The density contrast indicates the number density enhancement with respect to the average density in the vicinity of the galaxy (equation \ref{DC}). 

\begin{table*}

\caption{Density field measurements in the GOODS-S field (Full catalogs are published in the electronic edition)}
\centering 
\begin{tabular}{c c c c c c c} 
\hline\hline %inserts double horizontal lines
ID & RA(deg) & DEC(deg) & Redshift & \begin{tabular}{@{}c@{}}Comoving Density\\ $ h^{3} \text{Mpc}^{-3}$\end{tabular} & \begin{tabular}{@{}c@{}}Physical Density\\ $ h^{3} \text{Mpc}^{-3}$\end{tabular} & Density contrast \\ [0.5ex] % inserts table
%heading
\hline % inserts single horizontal line

  13889 & 53.1846685 &  -27.7875097  & 4.444  &   0.0112 &  1.7927 &  3.5429
  \\
  13890 & 53.1610507 &  -27.7883341 &  1.095  &   0.1478 &  1.3476 &  1.7321 
  \\
  13893 & 53.1540231 &  -27.7876987 &  1.782  &   0.0569 &  1.2397 &  1.0454   
  \\
  13894 & 53.021758  &  -27.7874826 &  4.878  &   0.0099 &  2.0489 &  7.0415   
  \\
  13895 & 53.1157124 &  -27.7876168 &  1.187  &   0.0564 &  0.6183 &  0.269    
  \\
  13896 & 53.0802905 &  -27.7874276 &  2.407  &   0.0194 &  0.7487 &  0.0688   
  \\
  13902&  53.1743111 &  -27.7876836 &  2.058  &   0.0223 &  0.6278  & -0.138 
  \\
  13903 & 53.0335614 &  -27.7880437 &  1.083  &   0.0781 &  0.6991  & 0.4059  
  \\
  13911 & 53.1471765 &  -27.7884795 &  0.631  &   0.1096  & 0.4745  & -0.2466  
  \\
  13912 & 53.1491844 &  -27.7878595 &  1.805  &   0.0553&   1.2311  & 0.9543  
\\
\hline %inserts single line
\end{tabular}
\label{table:examplecatalog}
\end{table*}

The comoving number density and density contrast of galaxies as a function of their redshifts are shown in figure \ref{ND}. The limiting magnitude of the survey restricts the sources to a certain stellar mass range. Hence, the evolution of the comoving number density with redshift is an inevitable result of missing low mass galaxies at higher redshifts.  In contrast, we find that the average density contrast is almost constant with redshift. This implies that the stellar mass function for a total sample of quiescent and star-forming galaxies does not change significantly with the environment. \citet{Davidzon16} have studied the effect of the environment on the shape of the galaxy stellar mass function up to redshift $z=0.9$, finding that the environmental dependence of the stellar mass function becomes weaker with redshift.

The histogram of the density contrast is also shown for each field in figure \ref{ND}. For all the fields, we find a similar distribution of density contrast, which has a dynamic range of $\sim 10$. For the entire sample of galaxies in all CANDELS fields, the average density contrast is 0.45 with a standard deviation of 0.75. It suggests that galaxies with a density contrast $\gtrsim 1.2$ are located in an over-dense region and those with density contrast $\lesssim -0.3$ reside in a void.  

Using the technique described in this paper, we estimate the density maps for all the five CANDELS fields. The evolution of the large scale structures is provided by 124 density maps covering $0.4 \leq z \leq 5$. The full density maps are available in the electronic version, with a few examples shown in the appendix (figure \ref{Density_map}). In the density maps, we limited the color-bar range to 5 to get a better contrast. Therefore, any density contrast above 5 is saturated with a dark red color.

\section{Results}
\label{sec:Result}
In this section, we use the estimated density fields to study the environmental effect on star formation activity of galaxies as a function of redshift. Here, we rely on the combined data from five widely separated fields to alleviate the cosmic variance effect as well as the sample size.    

\subsection{Environmental dependence of SFR and sSFR}

We investigate the dependence of SFR and sSFR on the local overdensity. We build a mass-complete sample of galaxies in four redshift intervals. Each interval contains galaxies with stellar mass greater than the completeness limit at that redshift. For example, the sample at $0.4 \leq z < 0.8$ consists of 6299 galaxies with $\rm M\geq M_{min}(0.8)$ where $\rm M_{min}(0.8)$ is the stellar mass completeness limit at $z=0.8$. The properties of the sample are summarized in Table \ref{Table_SFR_D}. Although we have density measurements up to $z=5$, we limit our investigation to $z\leq 3.5$. A mass-complete sample of galaxies at $3.5 < z \leq 5$, suffers from a small sample size ($<100$) and may not be used to draw any statistically significant conclusions.

\begin{figure*}[t]
    \centering
	\includegraphics[width=1\textwidth,clip=True]{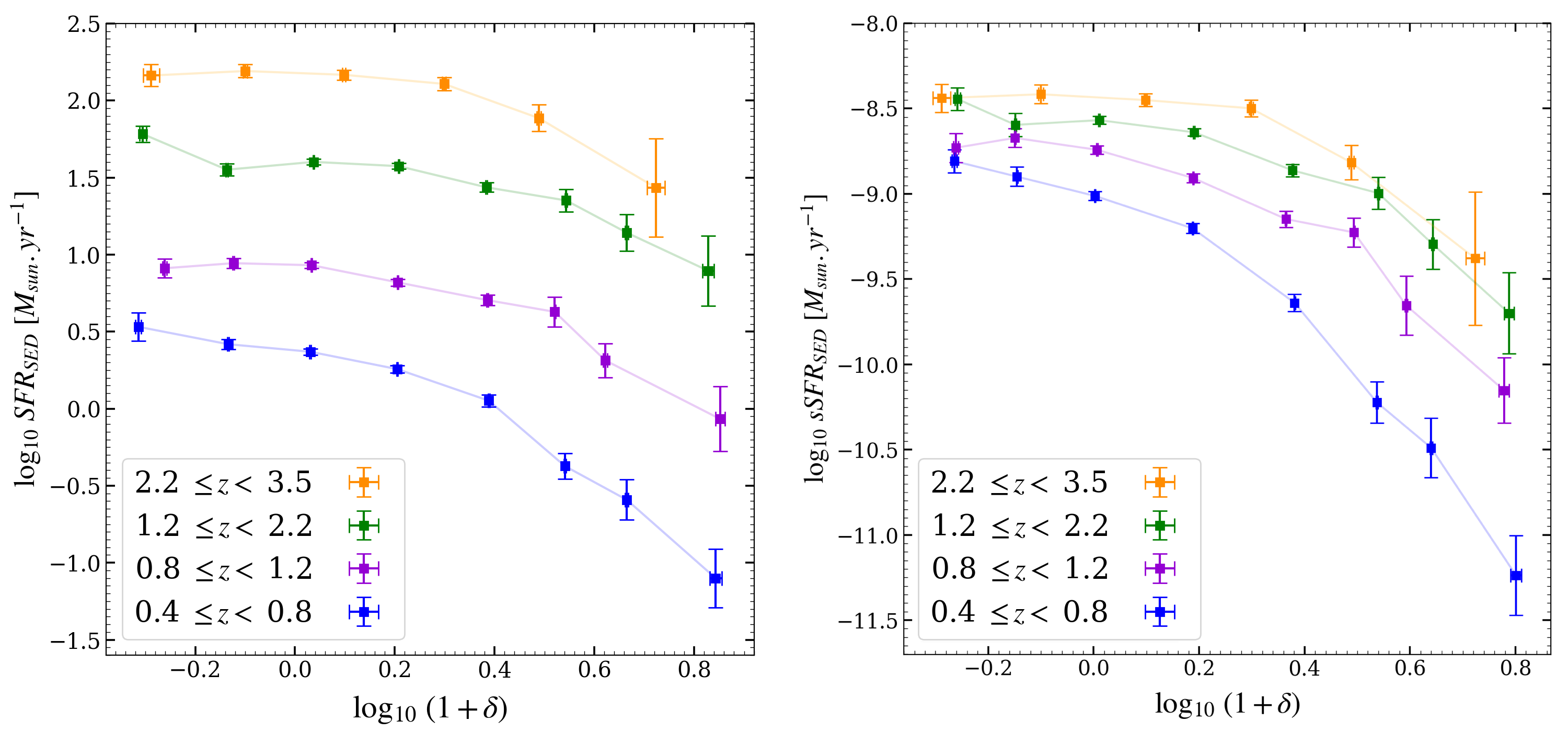}
	\caption{Environmental dependence of SFR and sSFR for a mass-complete sample of galaxies (Table \ref{Table_SFR_D}) at four redshift bins spanning across the redshift range of $z=0.4$ to $z=3.5$. The average SFR and sSFR of galaxies in density contrast bins are plotted as a function of overdensity ($1+\delta$). Error bars show the statistical uncertainty of the average values.}
	\label{SFR-D}
\end{figure*}

Figure \ref{SFR-D} demonstrates the average SFR and sSFR as a function of density contrast in the four redshift intervals. It shows a clear anti-correlation between sSFR and environmental density. The same trend can be seen in SFR-density relation.
At low redshift, $0.4\leq z < 0.8$, SFR decreases by a factor of $\sim 50$ as the density contrast increases from $\delta \sim -0.5$ to $\delta \sim 6$. This drop is steeper (by order of magnitude) for the sSFR. At $0.8\leq z < 1.2$, we find similar anti-correlation. These trends are in full agreement with previous studies \citep[e.g.,][]{Patel2009,Scoville13,Darvish16}. 

At high redshift, $1.2\leq z < 2.2$, we find that both SFR-density and sSFR-density relations follow the same trends we observe in the intermediate and low redshifts ($z<1.2$). The average SFR and sSFR of galaxies in dense environments are significantly lower compared to those residing in under-dense regions. For example, the average sSFR decreases $\sim 1.3$ dex with $\sim 1$ dex increase in density contrast. Several studies revealed the persistence of the environmental quenching at high redshifts out to $z\sim 2$ \citep[e.g.,][]{Grutzbauch12,Lin2012,Kawinwanichakij2017,Fossati17,Guo2017,Ji18}. Our results confirm that anti-correlation exists in both SFR-density and sSFR-density relations at least out to $z\sim 2.2$.           

Figure \ref{SFR-D} also shows that the environment does play a significant role at the highest redshift bin $2.2\leq z < 3.5$. SFR-density and sSFR-density relations behave in the same way that we observe in the local Universe. However, results need to be interpreted with caution since trends are found in different stellar mass ranges. For example, anti-correlation seen at $0.4\leq z < 0.8$ corresponds to the galaxies with $\log(\rm{M/M_{\odot}})>8.65$, while the relation at $2.2\leq z < 3.5$ includes only massive galaxies with $\log(\rm{M/M_{\odot}})>10.3$. The well-known relation between stellar mass and SFR in galaxies as a function of redshift confirms that the stellar mass plays a vital role in star formation activity of galaxies \citep[e.g.,][]{Peng10}. It suggests that star formation activity might be affected by both the local environment and stellar mass. Therefore, we disentangle the influence of environment and stellar mass in the next section before proceeding to the physical interpretation of the results.   
\begin{table}[h]

\caption{Properties of mass-complete sample}
\centering 
\begin{tabular}{ccc}

\hline\hline %inserts double horizontal lines

$\rm Redshift\ Range$ & $\ \ \  \log(\rm{M_{min}/M_{\odot}})\ \ \  $ & $\rm Sample\ Size$\\
%heading
\hline % inserts single horizontal line
\tableline 

$0.4\leq z < 0.8$ & 8.65& 6299  \\

 $0.8\leq z < 1.2$ & 8.98 & 6279  \\

$1.2\leq z < 2.2$ & 9.67 & 6168  \\

$2.2\leq z \leq 3.5$ & 10.30 & 1047  \\

\hline %inserts single line

\end{tabular}
\label{Table_SFR_D}
\end{table}

\subsection{SFR-Environment/Stellar Mass relation}\label{SFR-Environment/Stellar Mass relation}
\begin{figure*}
\centering  
\subfloat{\includegraphics[width=0.45\linewidth]{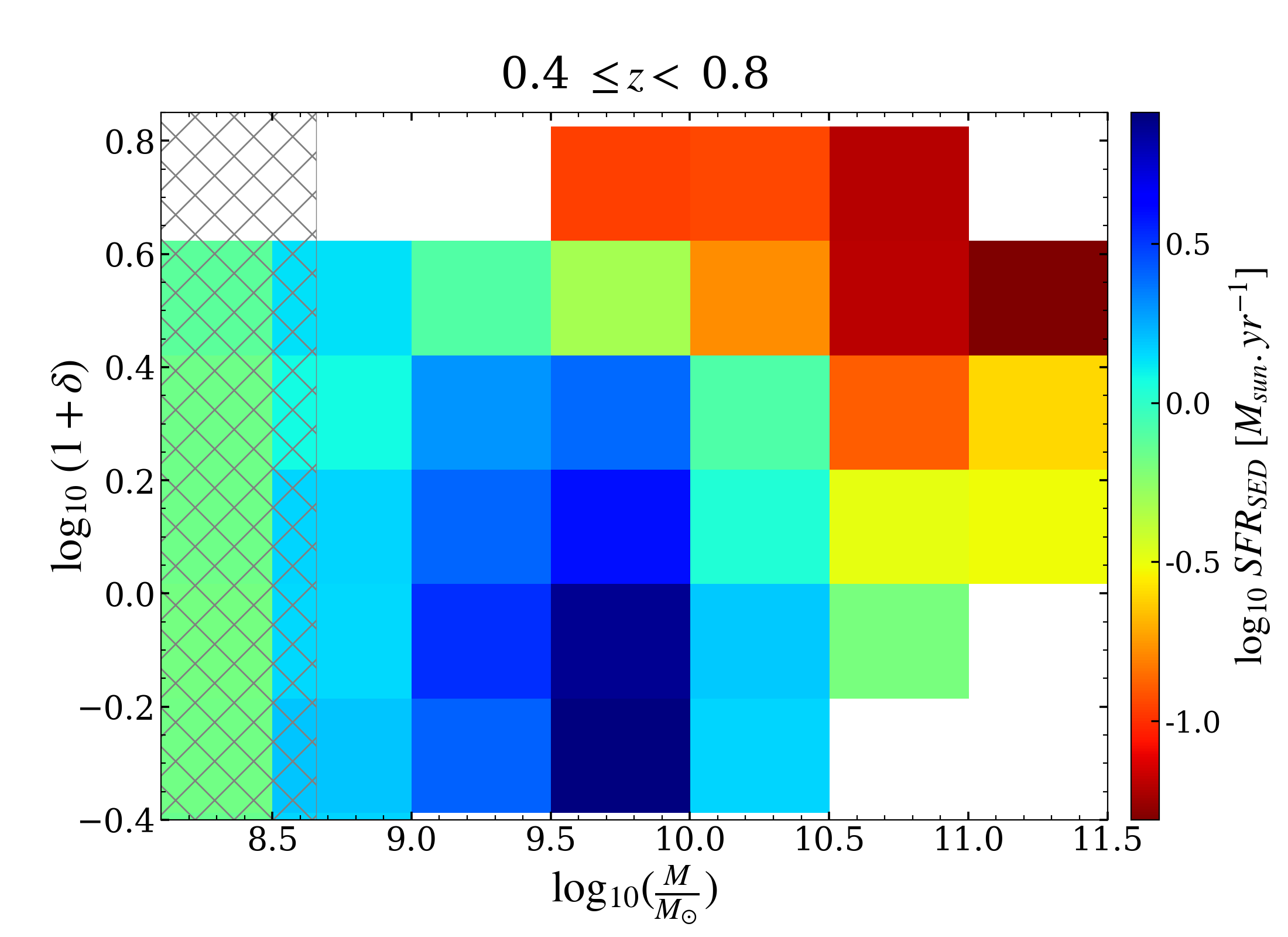}}
\qquad
\subfloat{\includegraphics[width=0.45\linewidth]{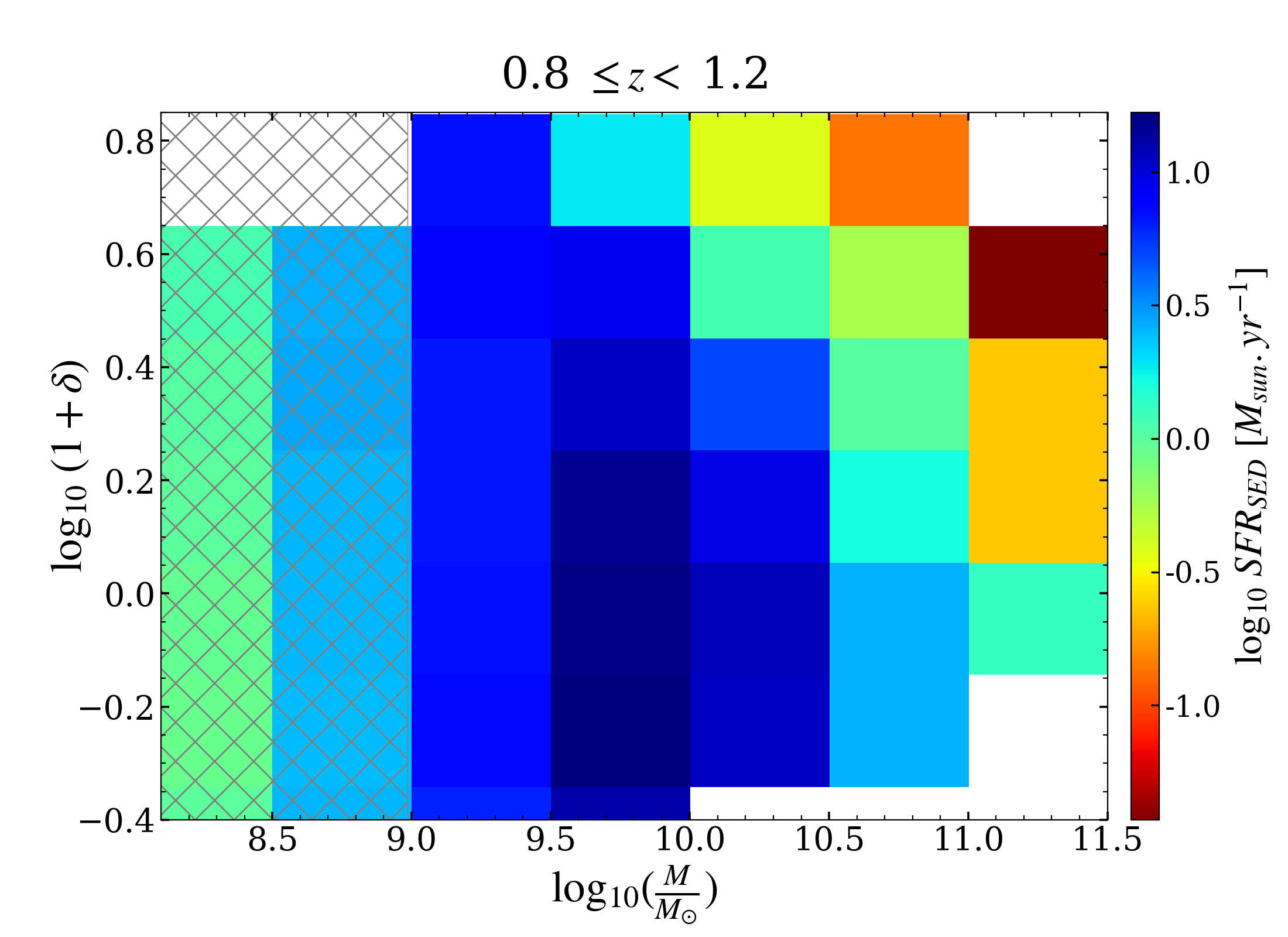}}
\qquad
\subfloat{\includegraphics[width=0.45\linewidth]{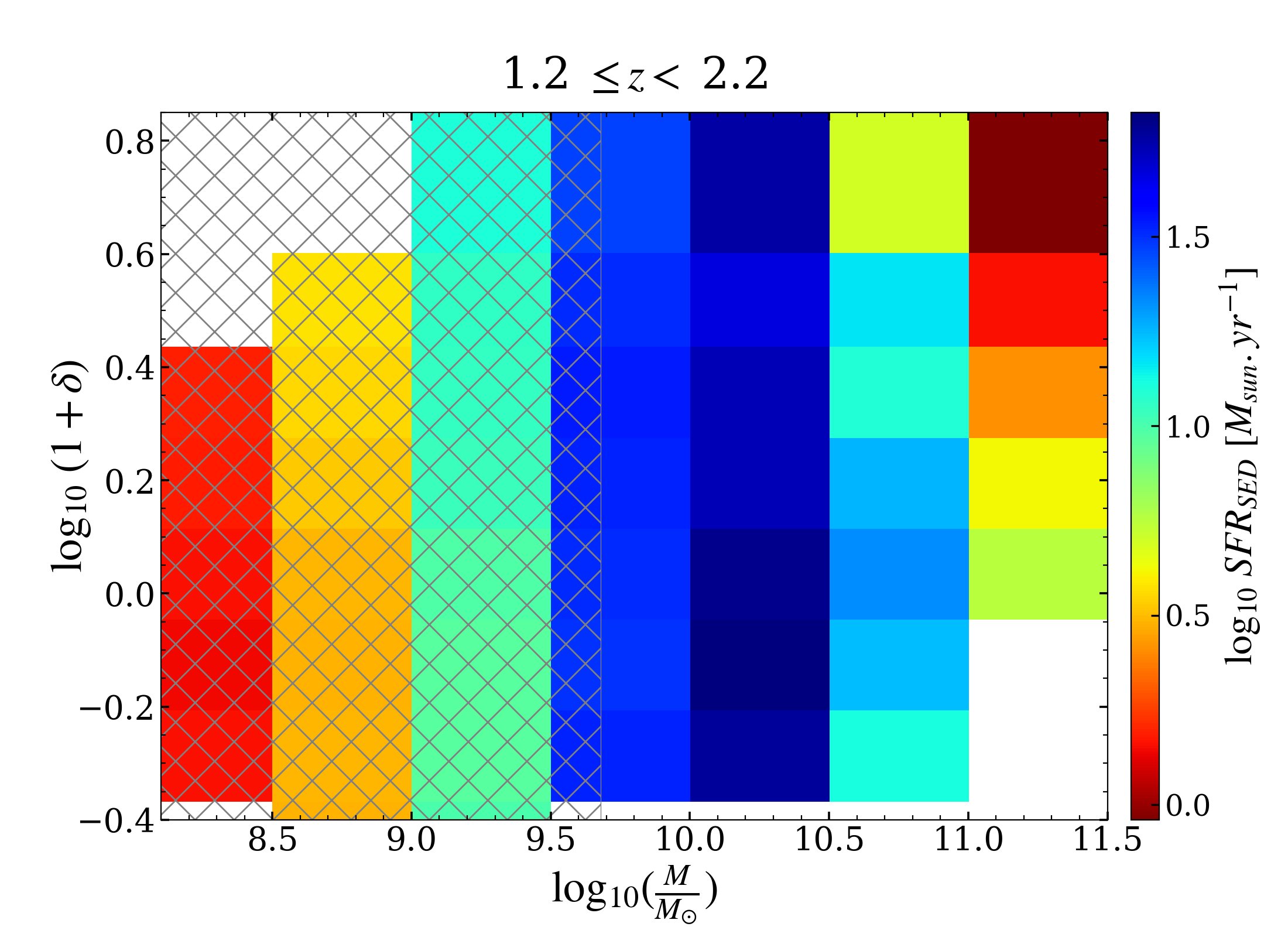}}
\qquad
\subfloat{\includegraphics[width=0.45\linewidth]{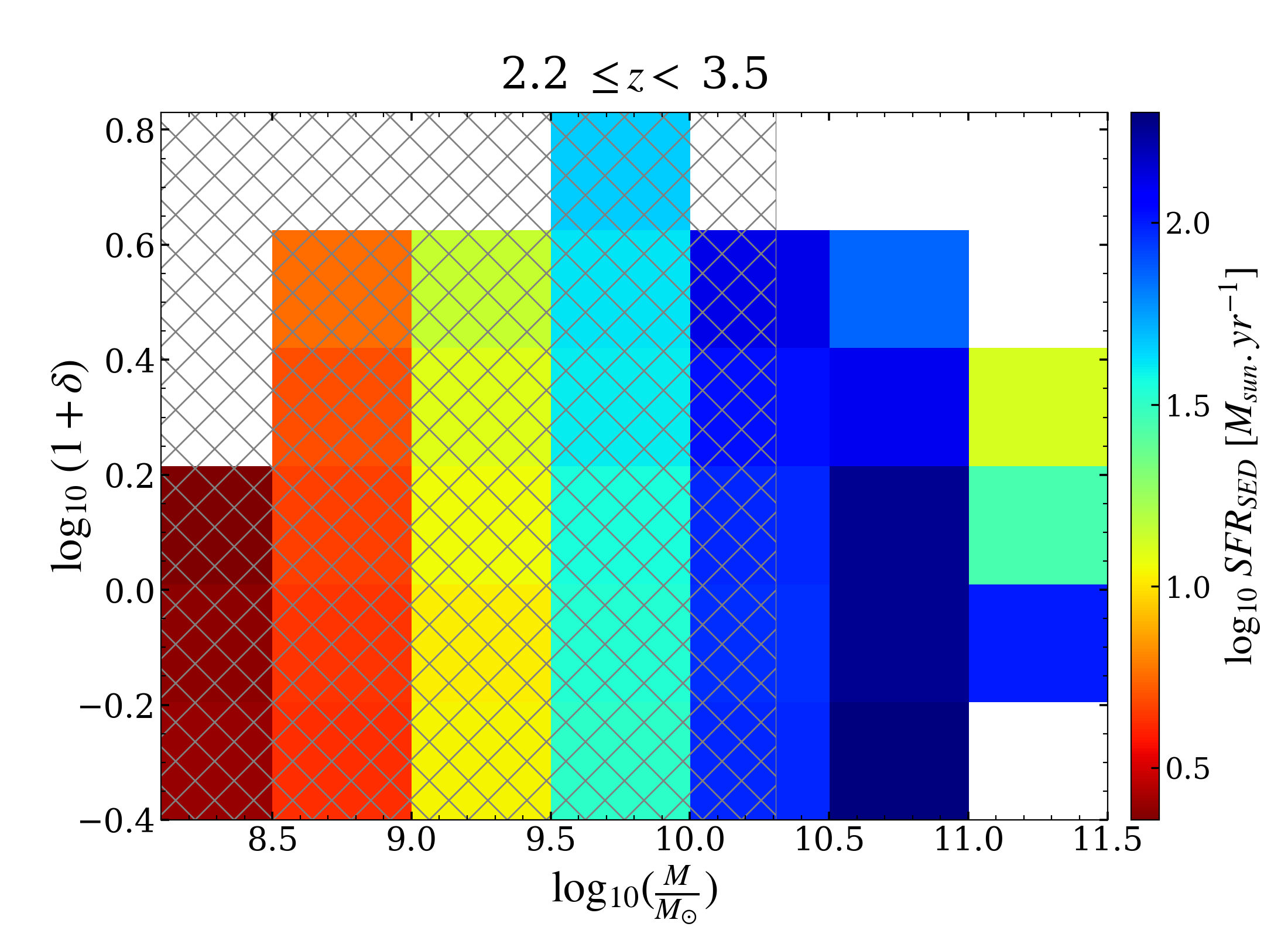}}
\caption{Average star formation rate in the bins of stellar mass and environment in four redshift intervals. The grey shaded regions show the incomplete stellar mass ranges. At all redshifts, we observe both environmental quenching and mass quenching for massive galaxies. We also find strong evidence of environmental quenching for low mass galaxies at low redshift.}
\label{SFR_delta_Mass}
\end{figure*}

Figure \ref{SFR_delta_Mass} presents the average SFR as a function of stellar mass and environment for the overall population of galaxies at the four redshift intervals. Colors indicate the average SFR in bins of environment and stellar mass. White areas show the regions with inadequate data points ($<20$).

We find that SFR of massive galaxies ($\rm M \geq 10^{11}\ M_{\odot}$) is inversely correlated with the environment at all redshifts ($0.4\leq z < 3.5$). For instance, at $1.2\leq z < 2.2$, massive galaxies in dense environments on average form their stars $\sim 6$ times slower than galaxies with the same stellar mass located in under-dense regions. In contrast, we do not find significant environmental dependence on SFR of galaxies with lower stellar masses ($\rm M < 10^{11}\ M_{\odot}$) at high redshifts ($1.2 \leq z < 3.5$). It reveals that the environmental quenching for very massive galaxies persists out to $z\sim 3.5$. This concurs well with the work done by \cite{Kawinwanichakij2017}, which is conducted out to $z=2$. Moreover, Figure \ref{SFR_delta_Mass} demonstrates strong evidence of environmental quenching for low mass galaxies ($\rm 10^{9.5}\ M_{\odot}<M< 10^{11}\ M_{\odot}$) at $z<1.2$ while it is not the case at higher redshifts ($z>1.2$).

We also investigate the fraction of quiescent galaxies as a function of stellar mass and environment. Similar to figure \ref{SFR_delta_Mass}, we find evidence of both stellar mass and environmental quenching out to $z\sim 3.5$, such that the fraction of quiescent galaxies increases with increasing density contrast and stellar mass. In order to quantify the efficiency of environment and stellar mass in galactic quenching, we adopt the method introduced by \citet{Peng10}. We define environmental quenching efficiency, $\varepsilon_{env}$, as the deficiency in the fraction of star-forming galaxies in the environment with overdensity $\delta$ compared to the under-dense region,   
\begin{equation}
    \varepsilon_{env}(\delta,\delta_0,\rm{M},z)=1-\frac{f_s(\delta,\rm{M},z)}{f_s(\delta_0,\rm{M},z)}
\end{equation}
where $f_s(\delta,\rm{M},z)$ is the fraction of star-forming galaxies with stellar mass M that are located in an overdensity $\delta$. $\delta_0$ is the density contrast of the under-dense environment. Following \cite{Kawinwanichakij2017} we consider the lowest 25 percentile of the $\delta$ distribution ($\delta_{25}$) as an under-dense environment ($\delta_0$) and we compute environmental quenching efficiency for galaxies that are located in an overdensity with $\delta$ greater than the 75 percentile of the $\delta$ distribution ($\delta_{75}$).  
A similar quantity can be defined for mass quenching efficiency, $\varepsilon_{mass}$,
\begin{equation}
    \varepsilon_{mass}(\delta,\rm{M},\rm{M_0},z)=1-\frac{f_s(\delta,\rm{M},z)}{f_s(\delta,\rm{M_0},z)}
\end{equation}

where $\rm M_0$ is the lowest stellar mass at any given redshift ($z$), which can be obtained by the stellar mass completeness limit ($\rm{M_{min}}(z)$). We compute mass quenching efficiency for galaxies with $\delta<\delta_{75}$.  

In order to calculate the fraction of star-forming galaxies, we separate star forming and quiescent galaxies based on their rest-frame U,V and J colors along with the \citet{UVJ} criteria. 

\begin{figure*}
    \centering
	\includegraphics[width=1\textwidth,clip=True, trim=3.2cm 0cm 5cm 2.2cm]{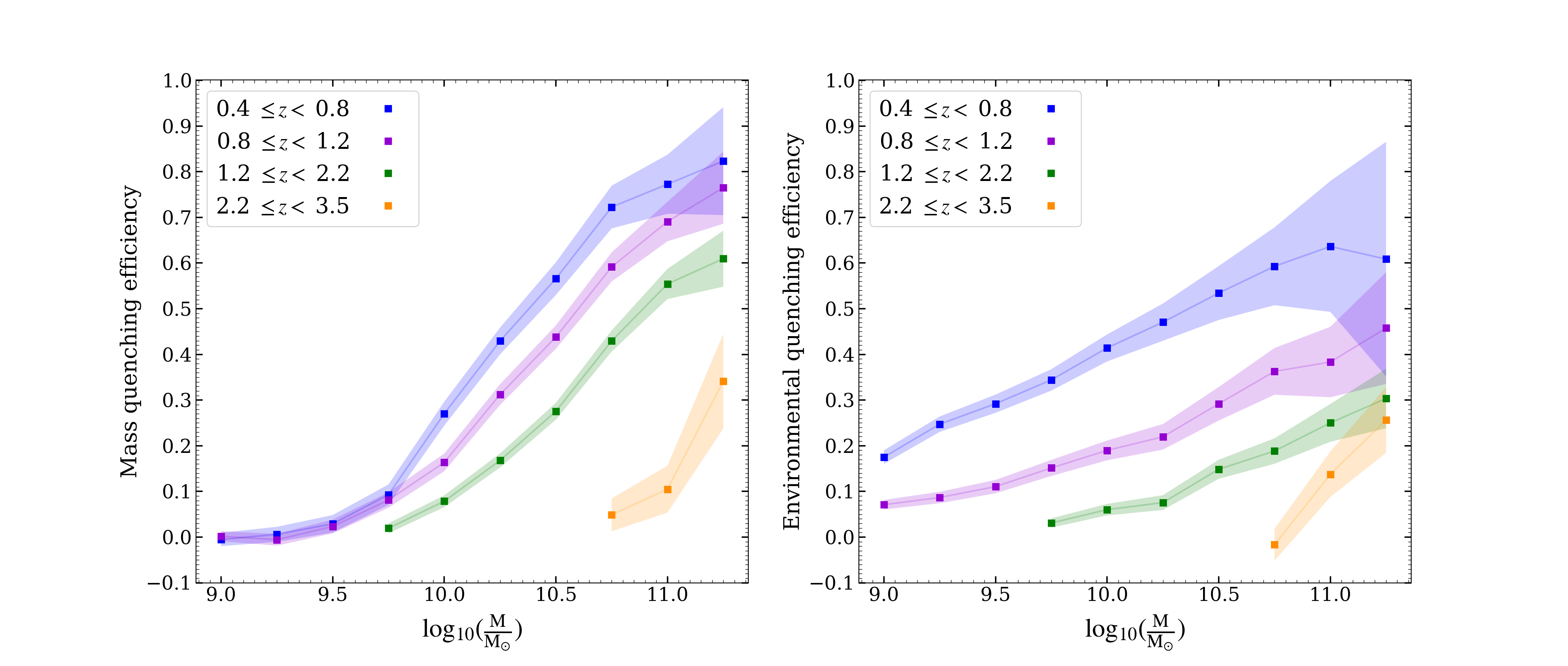}
	\caption{The mass quenching efficiency and environmental quenching efficiency, $\varepsilon_{env}(\delta>\delta_{75},\delta<\delta_{25},\rm{M})$ as a function of stellar mass. The  efficiencies are calculated in stellar mass bins, $\rm \Delta M \sim 0.5$ dex. Shaded regions show the uncertainty of efficiencies considering the Poisson statistics for the number of quiescent/star-forming galaxies.}\label{efficiency} 
\end{figure*}

The stellar mass dependence of the mass quenching efficiency, $\varepsilon_{mass}(\delta<\delta_{75},\rm{M},\rm{M_{min}}(z),z)$ and environmental quenching efficiency, $\varepsilon_{env}(\delta>\delta_{75},\delta<\delta_{25},\rm{M},z)$ are shown in Figure \ref{efficiency}. The efficiencies are calculated in stellar mass bins, $\Delta\rm{M}\sim 0.5$ dex, and error bars (shaded regions) are obtained considering the Poisson statistics for the number of quiescent/star-forming galaxies. 

At all redshifts, mass quenching efficiency increases significantly with stellar mass, which is consistent with previous works \citep[e.g.,][]{Peng10}. We also find that the environmental quenching efficiency is not independent of the stellar mass and it clearly increases with stellar mass, although this rise is weaker compared to the mass quenching efficiency \citep[see also][]{Lin,Papovich18}. At $z<1.2$, the environmental quenching is dominant for low mass galaxies ($\rm M \lesssim  10^{10} \rm {M}_\odot$). For example, at $0.4 \leq z <0.8$, the environmental quenching is $\sim 10$ times stronger than the mass quenching. For massive galaxies ($\rm M \gtrsim 10^{10} \rm {M}_\odot$), mass quenching is the dominant quenching mechanism at all redshifts; however, environmental quenching is significant for the most massive galaxies ($\rm M \gtrsim 10^{11} \rm {M}_\odot$). For instance, at $2.2 \leq z <3.5$, the environment and the stellar mass are almost equally responsible for the quenching of very massive galaxies ($\varepsilon_{mass}\sim \varepsilon_{env}$). This result reinforces our previous findings in Figure \ref{SFR_delta_Mass} that the environmental quenching of very massive galaxies exists at least out to $z\sim 3.5$. It also confirms that the environmental quenching is efficient for low mass galaxies at low and intermediate redshifts ($z<1.2$) \citep[see also][]{Peng10,Quadri2012,Scoville13,Lin,Lee2015,Darvish16,Nantais2016,Kawinwanichakij2017,Guo2017,Fossati17,Ji18}.

\subsection{Origin of the environmental quenching}
Although most of the studies, including this work, found strong evidence of the environmental quenching out to high redshifts, the physical mechanisms that are responsible are not clearly understood. \cite{vandeVooet2017} found a suppression of the cool gas accretion rate in dense environment at all redshifts, which becomes stronger at lower redshifts. This implies that a dense environment prevents the accretion of cold gas into the galaxy (cosmological starvation). As a result, the galaxy starts to consume the remaining gas reservoir in the depletion time scale, $t_{\rm{depl}}\propto \rm{M_{gas}}/\rm{SFR}$. This scenario is known as "over-consumption" model \citep{McGee2014,Balogh2016} and implies that the depletion time ($t_{\rm{depl}}$) depends on both stellar mass and redshift. The model predicts a short depletion timescale ($<100\ \rm Myrs$) for massive galaxies at high redshift. Therefore, "over-consumption" scenario could explain the environmental quenching that we observe here for massive galaxies at high redshifts \citep{Kawinwanichakij2017}. \citet{Feldmann2014} also showed through their simulation that, at $z\sim 3.5$, sSFR of a massive galaxy ($\rm M \sim  10^{11} \rm {M}_\odot$) drops by almost an order of magnitude within a few 100 Myrs. They found that this sudden halt at $z\sim 3.5$ is not caused by feedback processes and happens primarily due to the termination of the cool gas accretion. This provides another support that massive galaxies become quenched abruptly when their fresh gas accretion is terminated, possibly by locating in a dense environment. In addition, the lack of environmental quenching of low mass galaxies at high redshift can be explained by their low SFR, which results in longer depletion time \citep{Balogh2016,Kawinwanichakij2017}. 

Furthermore, "over-consumption" model could explicate our observations at low redshift (down to $z\sim 0.4$). For galaxies with stellar mass $\rm M \sim 10^{10.5} \rm {M}_\odot$, the depletion time ($t_{\rm{depl}}$) increases with decreasing redshift and reaches $\sim 2\ \rm{Gyrs}$ at $z\sim 0.4$ \citep{McGee2014} which is shorter than the typical dynamical time scale at that redshift ($t_{\rm{dyn}}\sim 4\ \rm{Gyrs}$) \citep{Balogh2016,Foltaz18}. It implies that the dynamical gas stripping processes are not required to explain our observation at low redshift. Moreover, the evolution of environmental quenching efficiency with stellar mass supports "over-consumption" model where the depletion time is longer for low mass galaxies resulting in weaker quenching efficiency. Therefore, "over-consumption" picture is most likely the dominant mechanism of environmental quenching, at least in the redshift range of this study. However, it is worth highlighting that at local Universe, the depletion time grows fast, such that it reaches $>10\ \rm{Gyrs}$ at $z=0$ for galaxies with intermediate stellar masses ($\rm M \sim 10^{10.5} \rm {M}_\odot$). This is longer than the dynamical time scale \citep{Balogh2016}. Consequently, "over-consumption" is likely not an effective quenching pathway at the local Universe and other dynamical processes are needed to explain the strong environmental quenching observed at $z\sim 0$ \citep{Peng10}.

\section{Discussion}
\label{sec:Discussion}
In this work, we introduce a robust method for reconstructing the underlying number density field of galaxies. The performance of KDE has been well explored by statisticians. They found that KDE can precisely estimate underlying densities of any shape, provided that the bandwidth is selected appropriately \citep[e.g.,][]{Silverman86}. We adopt a well-known Likelihood cross-validation (LCV) method to find the optimized bandwidth \citep[e.g.,][]{Hall}. Alternatively, one can use least squares cross-validation (LSCV) \citep{Bowman}, which is based on minimizing the integrated square error between the estimated and true densities. The LSCV method of bandwidth selection suffers the disadvantage of high variability \citep{Jones1996} and a tendency to under-smooth \citep{chiu1991}. We also correct densities for a systematic bias (under-estimation) near the edge of the survey using re-normalization. This assumes a symmetric galaxy distribution with respect to boundaries near the edges and may cause misestimated densities. This inevitable issue can be eliminated by observing as deep as CANDELS in a wider area.

\citet{Fossati17} have measured the environmental density for a $\rm JH_{140}\leq 24$ sample of 18,745 galaxies in the 3D-HST survey \citep{Skelton14} from $z=0.5$ to $3$, adopting circular aperture method (aperture radius fixed at 0.75 Mpc and width of z-slices at $\rm{\Delta}v=1500\ km.s^{-1}$). We find a significant difference between their density contrasts and our measurements. \citet{Fossati17} did not use the uniformly calculated photometric redshifts probability distributions. Instead, they assign redshifts based on the nearby galaxy with a spectroscopic redshift. They also use wider field public data for edge correction. We adopt the re-normalization method for edge-correction since we use the widest homogeneous fields with a depth of $\rm F160w=26\ AB\ mag$.

We explore any trends that may exist between the estimated densities and redshift in Figure \ref{ND}. This assures that the average density contrast does not evolve strongly with redshift. Otherwise, the diagram of any physical parameter (e.g., SFR, sSFR and quiescent fraction) as a function of density contrast would not be informative about the role of environment and trends could be affected by redshift evolution of physical parameters. 

Furthermore, the possibility of unrealistic trends due to the different assumptions in SED fitting (e.g., star formation history) also needs to be investigated. As a test case, we repeat our analysis based on the \citet{Pacifici2012} SED fitting method, which provides a library of SEDs assuming star formation histories from a semi-analytical model. Although the trends in Figure \ref{SFR-D} are more sensitive to SED fitting priors, the stellar mass dependence of environmental quenching efficiency (Figure \ref{efficiency}) does not change with new measurements. This reassures that the present evidence of environmental quenching at high redshift and the evolution of environmental quenching efficiency with stellar mass are not affected by SED fitting priors (section \ref{sec:Data}), especially by exponentially declining star formation history. 

\section{Summary}
\label{sec:Summary}
In this work, we report measurements of the environment for a $\rm F160w\leq26 \ AB \ mag$ sample of 86,716 galaxies in the five CANDELS fields (GOODS-South, GOODS-North, COSMOS, EGS, UDS) at $0.4 \leq z \leq 5$. We introduce a new method, \textit{boundary-corrected weighted von Mises kernel density estimation}, to reconstruct the underlying density field of galaxies. We find the optimal bandwidth for the von Mises kernel function in 124 z-slices spanning from $z=0.4$ to $5$ using the Likelihood Cross-Validation method. It allows us to create density field maps with the lowest bias/variance.   

We then use the density measurements to investigate the role of environment in star formation activity of a mass-complete sample of galaxies at $0.4\leq z\leq 3.5$. Our findings are summarized as follows: 
\begin{itemize}

\item[1-] At all redshifts, the average SFR and sSFR for a mass-complete sample of galaxies decrease with increasing density contrast. The trend is steeper at low redshift ($0.4\leq z<0.8$) such that the average SFR decreases by a factor of $\sim 50$ as the density contrast increases from $\delta \sim -0.5$ to $\delta \sim 6$.

\item[2-] We find strong evidence of environmental quenching for massive galaxies ($\rm M \gtrsim 10^{11} \rm {M}_\odot$) out to $z\sim 3.5$. We measured that the environmental quenching efficiency is $\gtrsim 0.2$, implying that a dense environment has $\gtrsim 20\%$ more massive quiescent galaxies compared to an under-dense region. This ratio reaches $\sim 60\%$ at the lowest redshift bin of this study ($0.4\leq z<0.8$).    
\item[3-] We find that the environmental quenching efficiency increases with stellar mass. This observation supports "over-consumption" model for environmental quenching where the gas depletion happens once the fresh gas accretion stops due to a dense environment. The gas depletion time depends on stellar mass and redshift and could explain the stellar mass dependence of the environmental quenching efficiency. The depletion time  becomes longer ($\rm > 10\ Gyr$) at lower redshifts, so it could not be a proper quenching pathway at local Universe; however, "over-consumption" is most likely the dominant environmental quenching mechanism at the redshift range of this study.

\end{itemize}

\section{ACKNOWLEDGEMENTS}

We thank the anonymous referee for providing insightful comments and suggestions that improved the quality of this work. We would like to acknowledge the contribution of Adriano Fontana, Janine Pforr, Mara Salvato, Tommy Wiklind, and Stijn Wuyts to the updated version of the photometric redshift catalog. NC would like to thank Ali Ahmad Khostovan for useful suggestions and discussions. 

\bibliography{environment_CANDELS}

\appendix
\counterwithin{figure}{section}
\section{Density maps}
We release over-density maps of 124 z-slices ranging from $z=0.4$ to $5$ for all CANDELS fields. A few examples are provided in figure \ref{Density_map}, but the full set of plots for the 124 z-slices are available in animation in the electronic version. In density maps, the color-bar range is limited to 5 to get a better contrast. As we expect, structures on the fields with a higher declination (e.g., GOODS-N and EGS) are elongated along the right ascension axis, which is the natural effect of mapping on (RA,DEC) coordinates. It should be recalled that the density contrast of a galaxy is inferred from multiple density maps (z-slices) considering its contribution in each z-slice, which is determined by photo-z PDF.    
\bigskip 
\begin{figure*}[h]
\centering  
\subfloat[COSMOS field]{\includegraphics[width=1\linewidth, clip=True, trim=0cm 0cm 0cm 0cm]{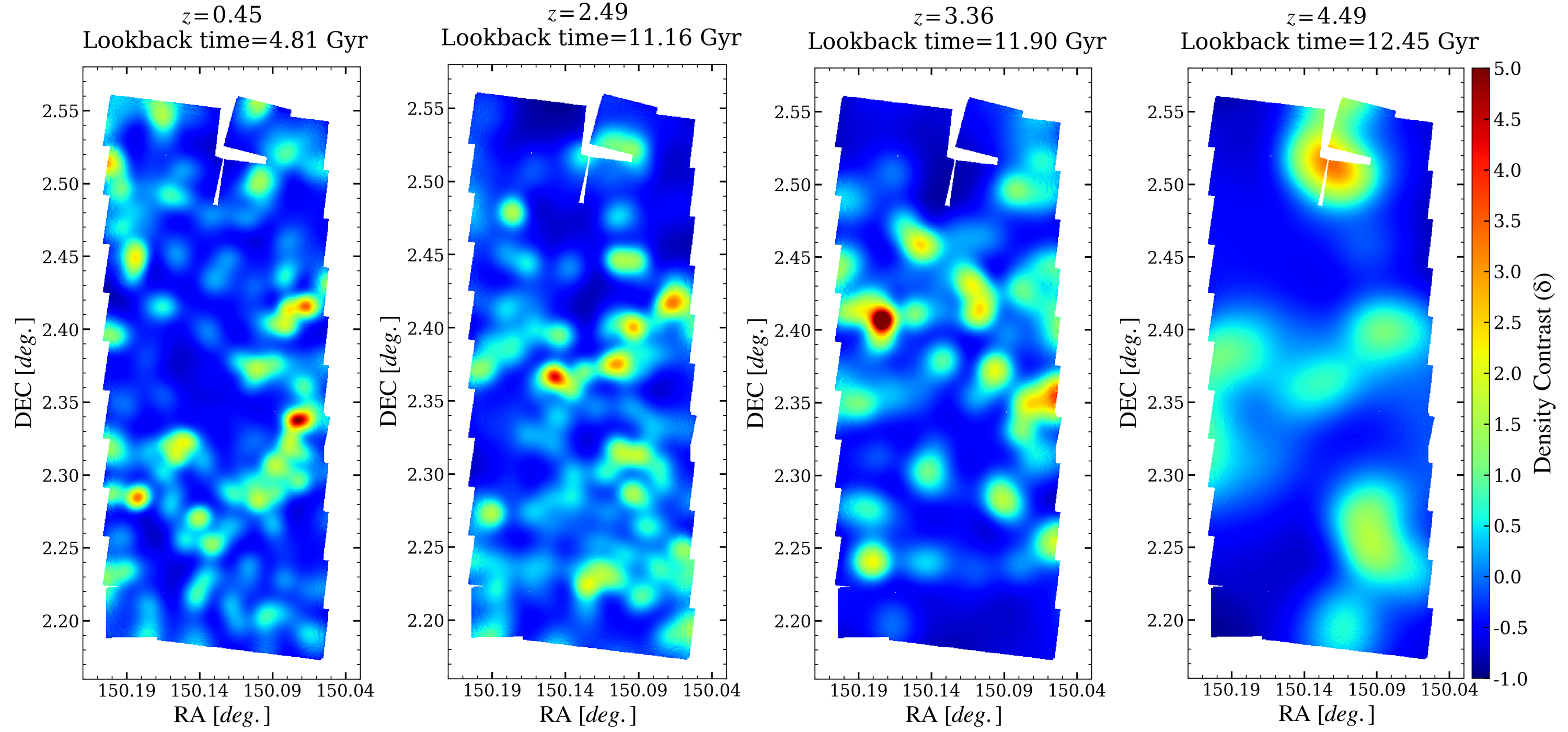}}
\qquad
\subfloat[GOODS-S field]{\includegraphics[width=1\linewidth, clip=True, trim=0cm 2cm 0cm 2cm]{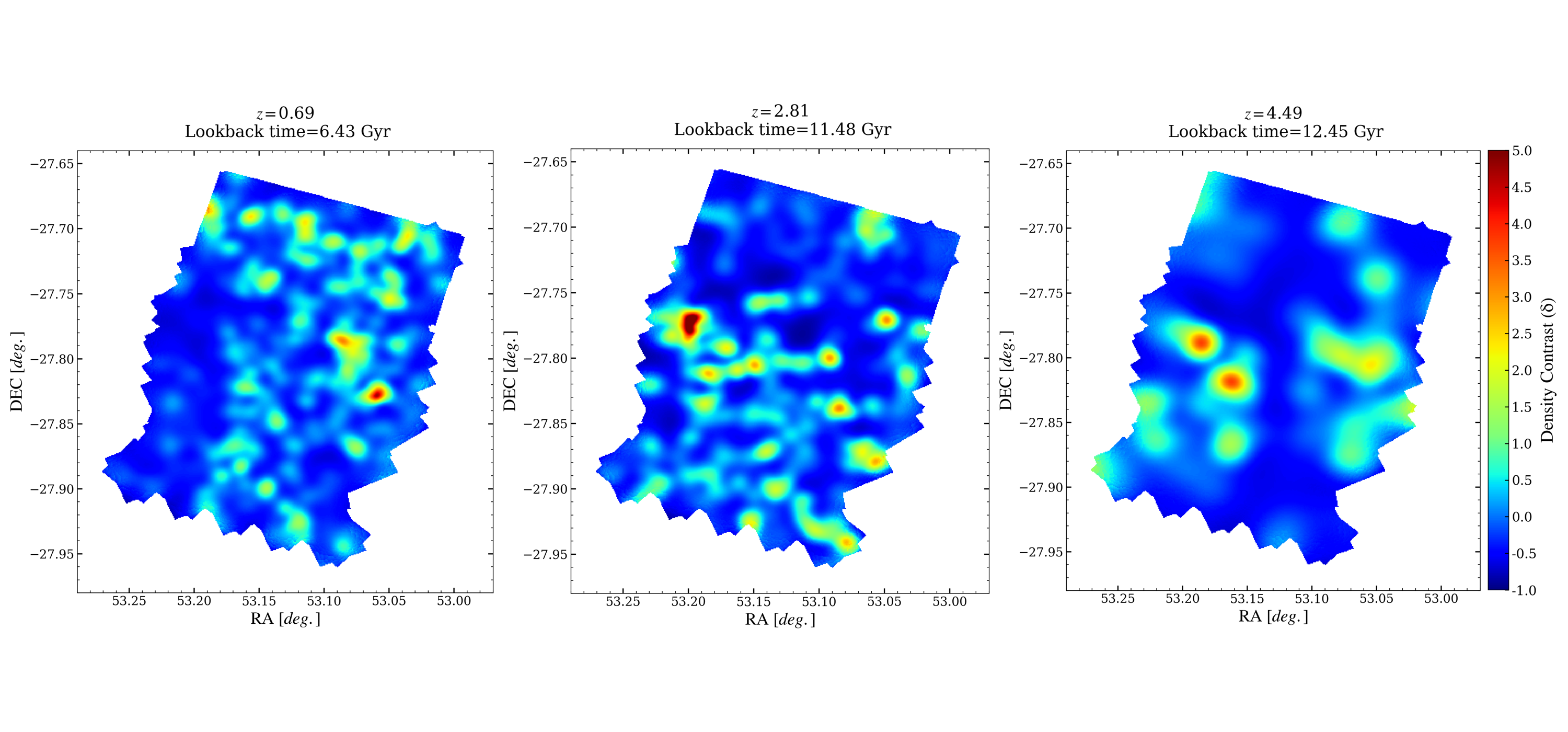}}

\caption{Density maps, plots for the 124 redshift slices are available in an animation format in the online version.}
\label{Density_map}
\end{figure*}

\begin{figure*}[ht]
\ContinuedFloat

\centering 

\subfloat[UDS field]{\includegraphics[width=1\linewidth, clip=True, trim=0cm 5cm 0cm 3cm]{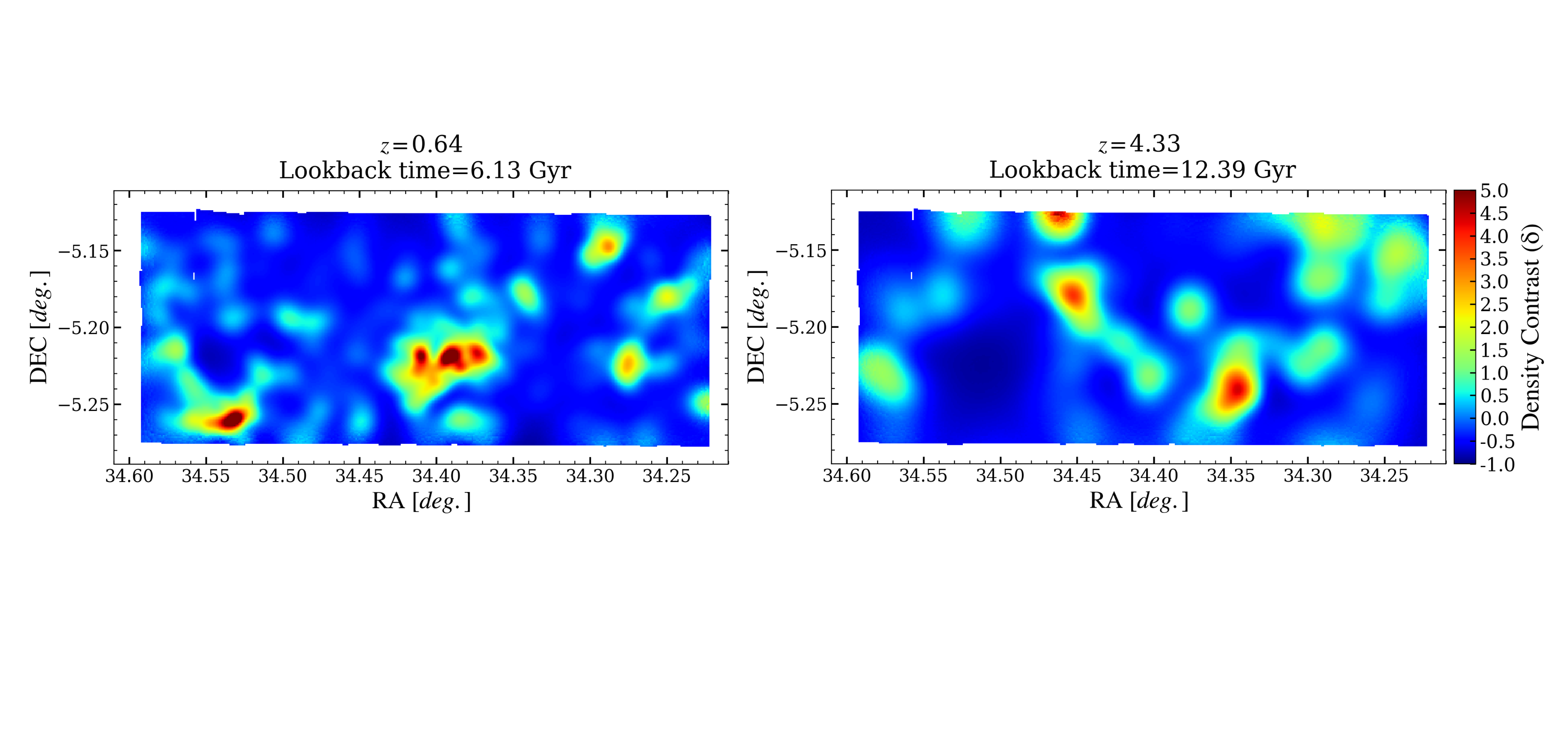}}
\qquad
\subfloat[GOODS-N field]{\includegraphics[width=1\linewidth, clip=True, trim=0cm 4cm 0cm 2.2cm]{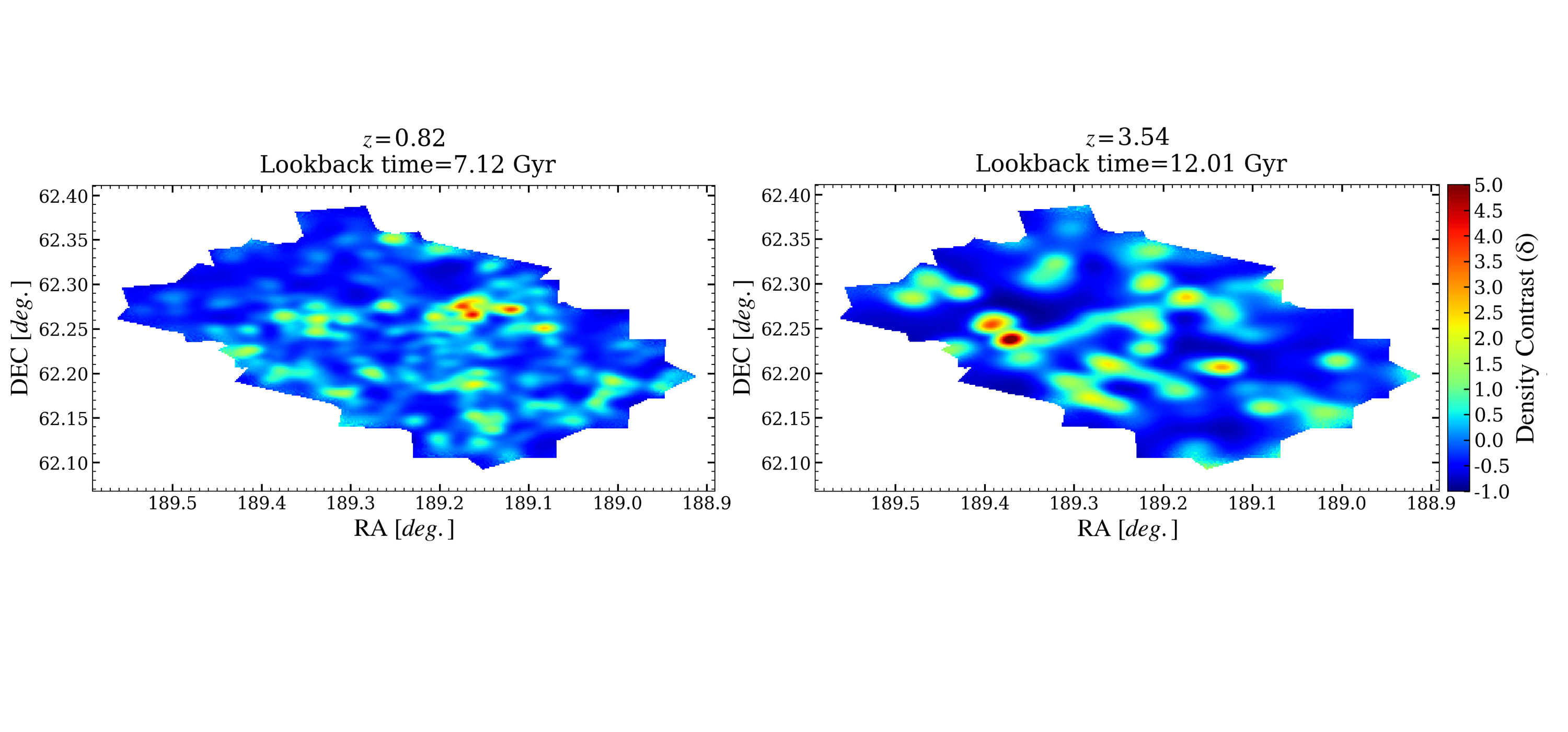}}
\qquad
\subfloat[EGS field]{\includegraphics[width=1\linewidth, clip=True, trim=0cm 2.2cm 0cm 2.2cm]{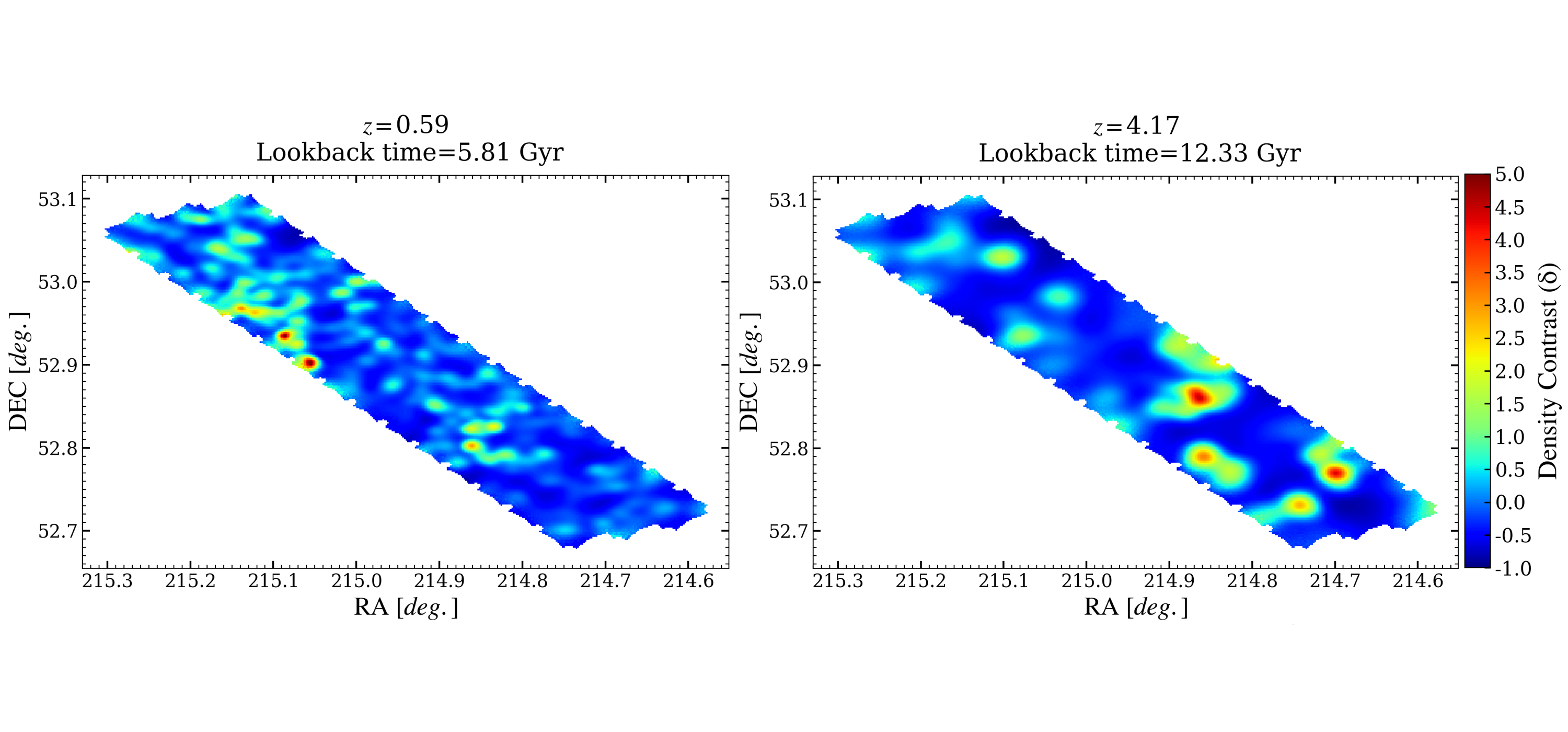}}

\caption{Continued: Density maps, plots for the 124 redshift slices are available in an animation format in the online version.}
\end{figure*}

\section{Redshift evolution of density measurements}
\begin{figure*}[hb]
    \centering
	\includegraphics[width=0.94\textwidth,clip=True, trim=3.4cm 4.5cm 5cm 6cm]{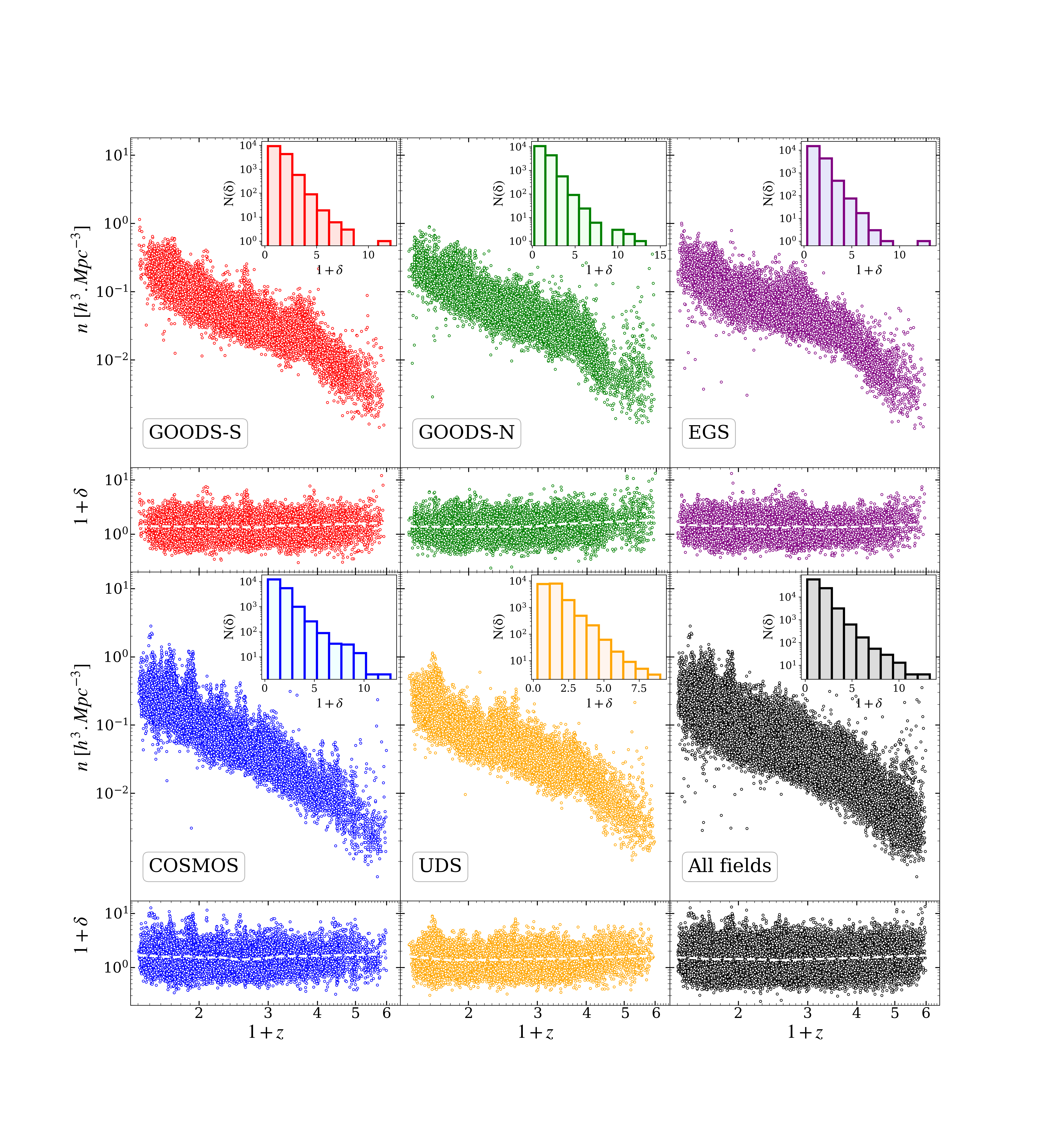}
	\caption{The comoving number density and the density contrast as a function of redshift as well as the density contrast histogram for each field. The comoving number density decreases with redshift due to the magnitude limit of the survey, while the average density contrast (white dashed lines) is almost constant over the cosmic time. This can be explained by the weak dependence of the stellar mass function on the environment. For all the fields, we find a similar distribution of density contrast, which has a dynamic range of $\sim 10$.}\label{ND} 
\end{figure*}

In this section, we investigate the correlation between density measurements and redshift. Figure \ref{ND} shows comoving number density and the density contrast as a function of redshift along with the distribution of density contrast separately for each field. Despite the clear evolution of comoving number density, the average density contrast, $\langle 1+\delta\rangle$, is almost constant over redshift. Although we find modest evidence of systematic trends between $\langle 1+\delta\rangle$ and redshift, especially at $z\gtrsim3$, the variation of $\langle 1+\delta\rangle$ over redshift is limited to $\lesssim 0.3$. Thus, the study of the physical properties of galaxies (e.g., SFR) versus density contrast (section \ref{sec:Result}) is not affected by the redshift evolution of overdensity measurements. We note that the average of $1+\delta$ is slightly higher than one since we do not define the density of the background ($\Bar{\sigma}$ in equation \ref{DC}) as the average density of galaxies. We define background density as the number of galaxies (computed from their photo-z PDFs) within each z-slice divided by the volume of that z-slice. 
\end{document}